\documentclass[%
 reprint,
showpacs,
amsmath,amssymb,
 aps,
prd,
]{revtex4-1}

\newcommand\ba{\begin{eqnarray}}
\newcommand\ea{\end{eqnarray}}
\newcommand\be{\begin{equation}}
\newcommand\ee{\end{equation}}

\usepackage{xcolor}

\usepackage[caption = false]{subfig}
\usepackage{graphicx}
\usepackage{dcolumn}
\usepackage{bm}

\usepackage{etoolbox}
\makeatletter
\patchcmd\linenumberpar{\@LN@parpgbrk}{\penalty\@LN@parpgpen\relax}{}{}
\makeatother

\begin{document}

\preprint{APS/123-QED}

\title{\boldmath{Study of the process $e^+ e^- \to p \bar p$ via initial state radiation at BESIII} \\}

\author{
\begin{small}
\begin{center}
M.~Ablikim$^{1}$, M.~N.~Achasov$^{10,d}$, P.~Adlarson$^{59}$, S. ~Ahmed$^{15}$, M.~Albrecht$^{4}$, M.~Alekseev$^{58A,58C}$, A.~Amoroso$^{58A,58C}$, F.~F.~An$^{1}$, Q.~An$^{55,43}$, Y.~Bai$^{42}$
% ...
\\
\vspace{0.2cm}
(BESIII Collaboration)\\
\vspace{0.2cm} {\it
$^{1}$ Institute of High Energy Physics, Beijing 100049, People's Republic of China\\
% ...
$^{4}$ Bochum Ruhr-University, D-44780 Bochum, Germany\\
% ...
$^{10}$ G.I. Budker Institute of Nuclear Physics SB RAS (BINP), Novosibirsk 630090, Russia\\
% ...
$^{15}$ Helmholtz Institute Mainz, Johann-Joachim-Becher-Weg 45, D-55099 Mainz, Germany\\
% ...
$^{42}$ Southeast University, Nanjing 211100, People's Republic of China\\
$^{43}$ State Key Laboratory of Particle Detection and Electronics, Beijing 100049, Hefei 230026, People's Republic of China\\
% ...
$^{55}$ University of Science and Technology of China, Hefei 230026, People's Republic of China\\
% ...
$^{58}$ (A)University of Turin, I-10125, Turin, Italy; (B)University of Eastern Piedmont, I-15121, Alessandria, Italy; (C)INFN, I-10125, Turin, Italy\\
$^{59}$ Uppsala University, Box 516, SE-75120 Uppsala, Sweden\\
% ...
$^{d}$ Also at the Novosibirsk State University, Novosibirsk, 630090, Russia\\
% ...
}\end{center}
\vspace{0.4cm}
\end{small}
}

\date{\today}
\begin{abstract}
The  Born cross section for the process $e^+ e^- \to p \bar p $ is measured  using  the initial state radiation technique with an undetected photon. This analysis is based on datasets corresponding to an integrated luminosity of 7.5 fb$^{-1}$, collected with the BESIII detector at the BEPCII collider at center of mass energies between 3.773 and 4.600~GeV. The Born cross section for the process $e^+ e^- \to p \bar p $  and the proton effective form factor are determined in the $p\bar p$ invariant mass range between 2.0 and 3.8~GeV/$c^2$ divided into 30 intervals. The proton form factor ratio ($|G_E|/|G_M|$) is measured in 3  intervals of the $p\bar p$ invariant mass between 2.0 and 3.0~GeV/$c^2$.

\end{abstract}

\pacs{13.66.Bc, 14.20.Dh, 13.40.Gp}
    \maketitle

\section{\label{sec:intro} Introduction}
Electromagnetic form factors (FFs) are fundamental quantities that describe the internal structure of  hadrons.  The proton (spin 1/2)  is characterized by the electric FF $G_E$ and the magnetic FF $G_M$.  They are experimentally accessible through the measurements of cross sections for elastic electron-proton scattering in the spacelike region (momentum transfer squared $q^2<0$) and annihilation processes $ e^+ e^- \leftrightarrow p \bar p $ in the timelike region ($q^2>0$) \protect\cite{Denig:2012by,Pacetti:2015iqa}.  At low momentum transfer,   spacelike FFs provide information on the distributions of the electric charges and magnetization within the proton.  In the timelike region, electromagnetic FFs can be associated with the time evolution of these distributions \cite{Kuraev:2011vq}.
  The unpolarized cross section for elastic electron-proton scattering has been measured for decades with improved accuracy. However, the recent data  on the elastic electron-proton scattering, based on the polarization transfer method \cite{Akhiezer:1968ek,Akhiezer:1974em}, showed that the ratio $\mu_p G_E/G_M$ (where $\mu_p$ is the proton magnetic moment) decreases almost linearly with $Q^2 = -q^2$ ~\cite{Puckett:2017flj}. This result is in disagreement with the previous measurements of unpolarized elastic $ep$ scattering~\cite{Puckett:2017flj}.
 
In the timelike  region, the proton FFs have been measured with the annihilation channels $ e^+ e^- \leftrightarrow p \bar p $ using the energy scan technique \protect\cite{Castellano:1973,Andreotti:2003bt,Ambrogiani:1999bh,Antonelli:1998fv,Bardin:1994am,Armstrong:1992wq,Delcourt:1979ed,Bisello:1983at,Bisello:1990rf,Ablikim:2005nn,Ablikim:2015vga,Pedlar:2005sj,Akhmetshin:2015ifg}, in which the center of mass (c.m.) energy ($\sqrt{s}$) of the collider is varied systematically, and at each c.m. energy point a measurement of the associated cross section is carried out.  The radiative return channel $e^+ e^- \to p \bar p  \gamma$, where $\gamma$ is a hard photon emitted by initial state radiation (ISR),  allows for a complementary approach to the energy scan technique in proton FF  measurements. It has been used by the {\it BABAR} Collaboration to measure the timelike proton FF ratio and the effective FF $|G_{\rm eff}(q^2)|$ (see Eq.~(\ref{eqEffdata}))  in a continuous range of $q^2$ \protect\cite{Lees:2013rqd,Lees:2013ebn}.   The {\it BABAR} data shows some oscillations in the measured $|G_{\rm eff}(q^2)|$. The origin of these oscillations has recently been the subject of several theoretical studies \cite{Lorenz:2015pba,Bianconi:2015vva}, but has not yet been well understood. The precision of the proton FF measurements in the timelike region has been limited by the statistics collected at the $e^+e^-$ and $p \bar p $ annihilation experiments.

In this paper we study the ISR process $e^+ e^- \to p \bar p  \gamma$ to measure the Born cross section of the process $e^+ e^- \to p \bar p $  and to determine  the proton FFs in the timelike region. We use datasets, corresponding to an integrated luminosity of 7.5 fb$^{-1}$, collected with the Beijing Spectrometer III (BESIII) \cite{Ablikim:2009aa} at the Beijing Electron-Positron Collider II (BEPCII) at c.m. energies between 3.773 and 4.600~GeV. We analyze the $e^+ e^- \to p \bar p  \gamma$ events in which the ISR photon cannot be detected because it is emitted at small polar angles (small-angle ISR), into the region not covered by the acceptance of the BESIII detector. The $e^+e^- \to p p\bar \gamma$ events are produced in the full range of the ISR polar angle. While only the final state proton and antiproton are detected, the small-angle ISR photon is identified based on the momentum conservation relations that describe this process. The differential cross section of the reaction $e^+ e^- \to p \bar p  \gamma$  as a function of the ISR polar angle reaches its highest values at small angles relative to the direction of the electron (or positron) beam \cite{Druzhinin:2011qd}. The measurement of the reaction $e^+ e^- \to p \bar p  \gamma$  in this region benefits from the availability of a large number of signal events.

\begin{figure}[h]
\includegraphics[height=3cm,width=5.5cm]{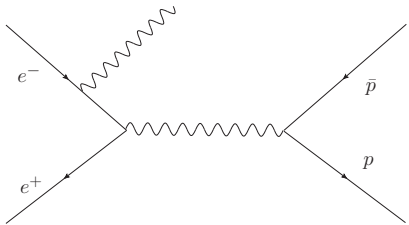}
\caption{Feynman diagram for the  ISR  process $e^+ e^- \to p \bar p  \gamma$. The ISR photon can be emitted from the electron or the positron.}
\label{isrfeyn}
\end{figure}

The Born cross section for the ISR process $e^+  e^-  \to p \bar  p \gamma$ (Fig.~\ref{isrfeyn}) integrated over the photon polar angle can be written as \cite{Druzhinin:2011qd}
\be
\frac{d\sigma_{e^+  e^-  \to p \bar p  \gamma}(q^2) }{dq^2}=\frac{1}{s} W(s,x) \sigma _{p \bar p }(q^2),
\label{eq:isrc2a}
\ee
where $q^2=M_{p\bar p }^2$, $M_{p\bar p }$ is the $p \bar p$ invariant mass, $x=\frac{2 E_{\gamma}^{*}}{\sqrt{s}}=1-\frac{q^2}{s}$, and $E_{\gamma}^{*}$ is the energy of the ISR photon in the $e^+ e^-$ c.m. system. The function \cite{Druzhinin:2011qd}
\be
W(s,x)=\frac{\alpha}{\pi x} \left(  \ln\frac{s}{m_e^2} -1 \right) (2-2x+x^2)
\label{eq:isrc2}
\ee
is the probability for the emission of a hard ISR photon with energy fraction $x$, $\alpha$ is the electromagnetic coupling constant, and $m_e$ is the electron mass.  Equations~(\ref{eq:isrc2a}) and (\ref{eq:isrc2})  describe ISR processes at the lowest QED order. The Born cross section for the nonradiative process $\sigma _{p\bar p }(q^2)$ is given by
\begin{equation} \label{eq:Borntot}
\begin{split}
\sigma _{p\bar p }(q^2)&=\frac{2 \pi \alpha^2 \beta C}{3 q^2 \tau}  \left( 2\tau|G_M|^2+|G_E|^2 \right),~ \tau=\frac{q^2}{4 M_{p}^{2}},\\
 \beta&=\sqrt{1-\frac{1}{\tau}},~C=\frac{y}{1-e^{-y}},~y=\frac{\alpha \pi}{\beta},
\end{split}
\end{equation}
where $M_p$ is the proton mass and $C$ is the Coulomb correction  factor \cite{Tzara:1970ne} which makes the cross section for the $p\bar p$ production nonzero at threshold.

The paper is organized as follows. The BESIII detector, the data and the Monte Carlo (MC) samples used in this analysis are described in Sec.~\ref{sec:detector}.  The procedure to identify the signal and to estimate the number of remaining background events is explained in Secs.~\ref{sec:selection} and \ref{sec:background}.  In Sec.~\ref{sec:totalxs} we present the results on the measurements of the Born cross section for the $e^+ e^- \to p\bar p $ channel and the proton effective FF. The measured values of the proton FF ratio and the branching fractions for the $J/\psi,~\psi(3686)$ to $p\bar p  $  decays are reported in Secs.~\ref{sec:FFratio} and \ref{sec:branching}, respectively.   The conclusion section contains a summary and an outlook.

\section{\label{sec:detector} The BESIII detector and event samples}
 BEPCII is a double ring $e^+ e^-$ collider running  at c.m. energies  between 2.0 and 4.6~GeV. It has a peak  luminosity of $ 1.0 \times 10^{33}$ cm$^{-2}$s$^{-1}$ at  $\sqrt{s}=3773$~MeV.  The  BESIII detector is a general purpose spectrometer with  an effective geometrical acceptance of $93\%$ of $4\pi$. It consists of a small cell, helium-based ($60\%$ He, $40\%$ C$_3$H$_8$) main drift chamber (MDC),   a time-of-flight  (TOF) system, a  CsI(Tl) electromagnetic calorimeter (EMC) and a muon system (MUC). The MDC provides momentum measurement of charged  particles with a resolution of $0.5\%$ at 1~GeV/$c$ in a 1 Tesla magnetic field. The energy loss measured by the MDC has a resolution better than $6\%$. The TOF is based  on  5-cm-thick plastic scintillators with a time resolution of 80 ps in the barrel and 110 ps in the end caps. The EMC  is used to measure the energies of photons and electrons. The  EMC provides an energy resolution  (for 1~GeV photons) of $2.5\%$ in the barrel region and $5.0\%$ in the end caps. The MUC system consists of resistive plate chambers. It is used to identify muons and provides a spatial resolution better than 2 cm.

\begin{table}
\caption{Integrated luminosities  of the data samples used in this analysis \cite{Ablikim:2015orh, Ablikim:2015nan}. The quoted uncertainties are statistical and systematic, respectively.}
\label{dataset}
\begin{ruledtabular}
\begin{tabular}{cc}
 $\sqrt{s}$ [GeV]   &    Integrated luminosity [pb$^{-1}$]   \\\hline
    3.773                  &   2931.8 $\pm$ 0.2  $\pm$ 13.8    \\
    4.008                  &   481.96 $\pm$ 0.01  $\pm$  4.68  \\
    4.226                  &   1053.9 $\pm$ 0.1   $\pm$ 7.0 \\
    4.258                  &   825.67 $\pm$  0.13  $\pm$  8.01  \\
    4.358                  &   539.84  $\pm$ 0.10  $\pm$ 5.24  \\
    4.416                  &   1041.3 $\pm$ 0.1  $\pm$ 6.9  \\
    4.600                  &   585.4  $\pm$ 0.1  $\pm$ 3.9 \\
\end{tabular}
\end{ruledtabular}
\end{table}

The data samples used in this analysis were collected at 7 c.m. energy points between 3.773 and 4.600~GeV. Table~\ref{dataset} summarizes the integrated luminosity collected at each c.m. energy point \cite{Ablikim:2015orh, Ablikim:2015nan}.  The integrated luminosities of the datasets used in this work were measured using the Bhabha scattering events.  Their systematic uncertainties are mainly due to the uncertainties on the tracking of charged particles, the estimation of the signal selection efficiency,  the determination of  the c.m. energy, and the trigger efficiency for collecting the  Bhabha scattering events in the online data acquisition.
 MC samples for signal and background channels are simulated using a {\sc geant4}-based \cite{Agostinelli:2002hh} simulation software package BESIII BOOST (BESIII Object Oriented Simulation Tool) \cite{Deng:2006}. The MC samples are produced with large amounts of generated events to determine the signal efficiencies and to estimate the potential background contamination.
The signal process  $e^+ e^- \to p\bar p  \gamma$ is generated with the {\sc phokhara}  event generator \cite{Czyz:2014sha}, which takes into account next-to-leading order radiative corrections. The critical background channels $e^+ e^- \to p\bar p  \pi^0 (\gamma)$ and the two-photon process ($e^+e^- \to e^+e^- f^+f^-$, where $f$ can be leptons, or quarks which hadronize using {\sc jetset} \cite{Sjostrand:1994ek}) are simulated using the generator software package {\sc conexc} \cite{Ping:2013jka}  and the event generator {\sc bestwogam} \cite{Ping:2008zz}, respectively.  The ISR background processes   $e^+ e^- \to \mu^+ \mu^- \gamma,~ \pi^+ \pi^- \gamma$ and $K^+ K^-  \gamma$ are simulated with the {\sc phokhara}   event generator  up to the next-to-leading order of radiative corrections.  The inclusive hadronic channels $e^+ e^- \to q \bar q  ~(q=u,~d,~s)$ are studied with the {\sc kkmc} event generator  \cite{Jadach:2000ir,Jadach:1999vf}. The  $e^+ e^- \to e^+ e^- \gamma$ channel is simulated with the {\sc babayaga} event generator \cite{Balossini:2006wc}.
The ISR processes  $e^+e^-\to\gamma J/\psi,~\gamma\psi(3686),~\gamma\psi(3773)$ and $\gamma\psi(4040)$ are generated with {\sc besevtgen} \cite{Ping:2008zz} using the {\sc vectorisr} model \cite{Bonneau:1971mk,Lange:2001uf}.

\section{\label{sec:selection} Event Selection}
Charged tracks  of polar angles $|\cos\theta|<0.93$  are identified by the MDC.  The distance between the  interaction point (IP) and the point of closest approach for each charged track  is required to be within 1 cm in the plane perpendicular to the beam direction and  within $\pm 10$~cm  along the beam  direction. The energy loss in the MDC and the flight time measured by the TOF system are used to calculate the  particle identification (PID) probabilities for the electron, muon, pion, kaon and proton hypotheses. The particle type of highest PID probability is assigned to the charged track.  The ratio of the shower energy deposited in the EMC  ($E_{\rm EMC}$) to the reconstructed momentum ($p_{\rm rec}$) of the positively charged track associated with the shower is required to be less than 0.5.  The PID efficiency for the proton and the antiproton, in the momentum range between 0.3 GeV/c and 1.5 GeV/c,  is larger than $90~\%$. The events with only two charged tracks, identified as proton and antiproton, are selected.

\begin{figure}[h]
\includegraphics[scale=0.45]{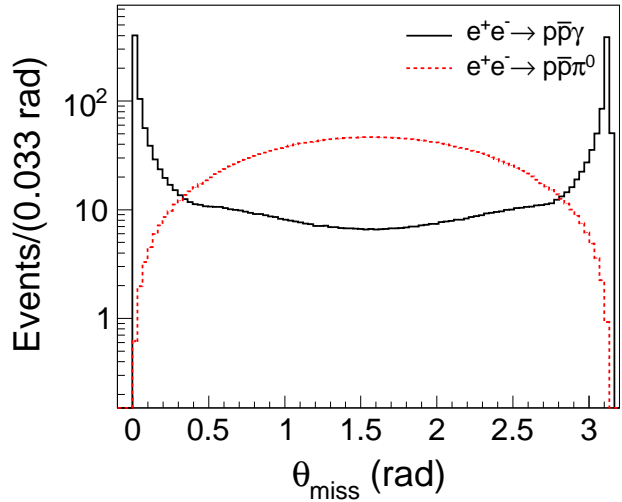}
\caption{Distributions of  $\theta_{\rm miss}$ for the simulated signal events  $e^+ e^- \to p \bar p  \gamma$ (black solid) and $e^+ e^- \to p \bar p  \pi^0$  (red dashed),   at $\sqrt{s}=4.226$~GeV. These distributions are normalized to the numbers of the expected events in the data sample according to their cross sections and luminosity.}
\label{misstheta}
\end{figure}

In this analysis, the ISR photon is not detected. The final event selection is  based  mainly on two variables, the missing momentum $\vec{p}_{\rm miss}$ and the missing mass squared $M_{\rm miss}^{2}$  recoiling against the $p\bar  p $ system. The missing momentum is defined as:
\be
\vec{p}_{\rm miss}= \vec{k}_{1}+\vec{k}_{2}-\vec{p}_{1}-\vec{p}_{2},
\ee
where $\vec{k}_1$ ($\vec{k}_2$) and $\vec{p}_1$ ($\vec{p}_2$) are the momentum vectors in the laboratory frame of the initial state electron  (positron) and final state antiproton (proton), respectively.   The angular distribution of the missing momentum is used to suppress the hadronic background, in particular the process $e^+ e^- \to p \bar p  \pi^0$.  Figure~\ref{misstheta} shows the distribution of the polar angle ($\theta_{\rm miss}$) of the missing momentum in the laboratory frame for the MC signal and  $e^+ e^- \to p \bar p  \pi^0$ background events. The angle $\theta_{\rm miss}$  is required to be in the region
\be
\theta_{\rm miss} <0.125 ~\mbox{or}~ \theta_{\rm miss}>(\pi-0.125)~\mbox{rad}.
\label{ctheta4230}
\ee
This condition removes the signal events in which the ISR photon is emitted at large polar angle.

\begin{figure}[h]
\includegraphics[height=7cm,width=9cm]{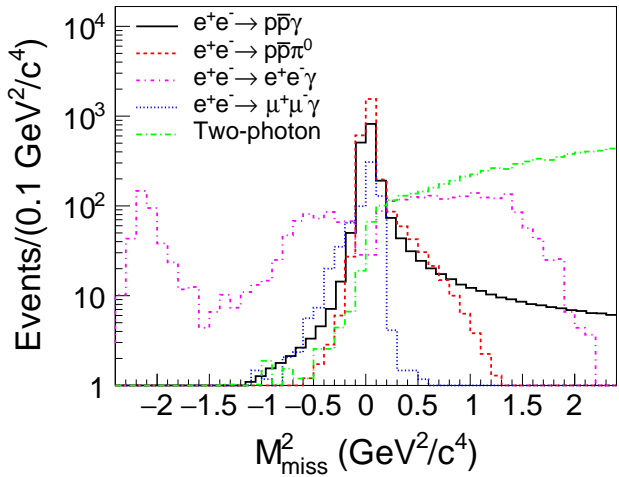}
\caption{Distributions of $M_{\rm miss}^2$  for the simulated signal events  $e^+ e^- \to p \bar p  \gamma$ (black solid), $e^+ e^- \to p \bar p  \pi^0$  (red dashed), $e^+ e^- \to  e^+ e^- \gamma$  MC (purple dashed-dotted), $e^+ e^- \to  \mu^+ \mu^- \gamma$  MC (blue dotted) and for the two-photon production (green long dashed-dotted) after charged track selection (before applying the  $\theta_{\rm miss}$ condition),  at $\sqrt{s}=4.226$~GeV.  These distributions are normalized to the numbers of the expected events in the data sample according to their cross sections and luminosity. The long positive tail in the  distribution of $M_{\rm miss}^2$  for the signal events is due to the extra photon emission which takes into account next-to-leading-order radiative corrections.}
\label{missmass3773}
\end{figure}

The missing mass squared is defined by

\be
M_{\rm miss}^2= (K_{1}+K_{2}-P_{1}-P_{2})^2,
\ee
where $K_1$ ($K_2$)  and $P_1$ ($P_2$) are  the four-momenta of the initial state electron (positron) and final state antiproton (proton), respectively. Figure~\ref{missmass3773} shows the distributions of $M_{\rm miss}^2$ for the simulated signal and background events at $\sqrt{s}=4.226$~GeV.  The events are required to have  a $M_{\rm miss}^2$ in the interval
\be
-0.1<M_{\rm miss}^2<0.2~\mbox{GeV$^2$/$c^4$},
\label{cmiss4230}
\ee
for the data samples collected  at $\sqrt{s}>4$~GeV, and
\be
-0.02<M_{\rm miss}^2<0.10~\mbox{GeV$^2$/$c^4$},
\label{cmiss3773}
\ee
for the data sample collected  at $\sqrt{s}=3.773$~GeV. 
This condition mainly suppresses the background from the processes $e^+e^-\to e^+e^-\gamma,~ e^+e^-\to p \bar p \pi^0 \gamma$, and two-photon channel.  At $\sqrt{s}=3.773$~GeV,  a narrower window of  the  $M_{\rm miss}^2$ interval is needed to reject the remaining background from the resonance [$J/\psi,~\psi(3686)$] decays into the $p \bar p \gamma$ final state.

The polar angles of the proton and the antiproton in the $p \bar p$ c.m. system are required to be   within $|\cos\theta_{p,\bar p}^{{p \bar p}}|<0.75$. Due to the conditions applied on the distributions of   $\theta_{\rm miss}$ and $M_{\rm miss}^{2}$  [Eqs.~(\ref{ctheta4230}],  (\ref{cmiss4230}) and (\ref{cmiss3773})), the efficiency of the signal in the region  $|\cos\theta_{p,\bar p}^{{p \bar p}}|>0.75$ is very small. The condition   $|\cos\theta_{p,\bar p}^{{p \bar p}}|<0.75$ is used to suppress the remaining background from the process  $e^+ e^- \to e^+ e^- \gamma$.

\begin{figure}[h]
\includegraphics[height=7cm,width=9cm]{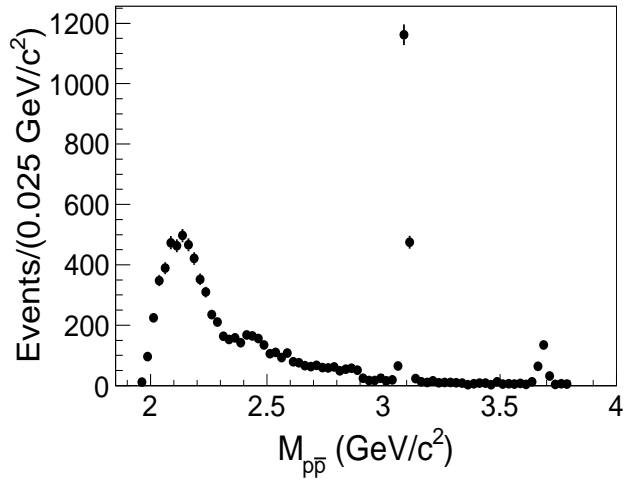}
\caption{The distribution of $M_{p\bar p }$ for the combined selected data events.}
\label{invmassdata}
\end{figure}

 The collected events at the 6 c.m. energies for $\sqrt{s}>4$~GeV are analyzed in $M_{p\bar p }$ intervals between 2.0 and 3.8~GeV/$c^2$. The events collected at  $\sqrt{s}=3.773$~GeV  are analyzed in a smaller  $M_{p\bar p }$  range  between 2.0 and 2.9~GeV/$c^2$. Above   2.9~GeV/$c^2$    (3.8~GeV/$c^2$), the number of signal events from  $\sqrt{s}=3.773$~GeV ($\sqrt{s}>4$~GeV) is  small and it is comparable to the number of remaining background events.  The distribution of $M_{p\bar p }$  for the selected data candidates is shown in Fig.~\ref{invmassdata}.   The total number of events,  from the data samples collected at the 7 c.m. energies,  is around 9100. Selected events from  $J/\psi \to p \bar p $ and $\psi(3686) \to p \bar p $ decays are clearly seen at $M_{p \bar p} \sim 3.1$ and 3.7~GeV/$c^2$,    respectively.

\section{\label{sec:background} Background estimation and subtraction}
The background events in the MC samples of  $e^+ e^- \to e^+ e^- \gamma,~ \mu^+ \mu^- \gamma,~ \pi^+ \pi^- \gamma$ and $K \bar K \gamma$  are suppressed by the selection criteria described in Sec.~\ref{sec:selection}. The amount of generated events in each MC sample exceeds the number of expected events for these background channels according to their cross sections and luminosities, and they can consequently be safely neglected.
  The ISR channels $e^+e^-\to\gamma {\cal{R}}~({\cal{R}} \to p \bar p  \gamma),~{\cal{R}}= J/\psi, \psi(3686), \psi(3773), \psi(4040)$ are suppressed to below $0.5\%$ of the total selected events and they can also be neglected. In the following the numbers of background events from  $e^+e^-\to\gamma {\cal{R}}~({\cal{R}} \to p \bar p ),~{\cal{R}}= J/\psi, \psi(3686)$, $e^+ e^- \to p \bar p  \pi^0$  and the two-photon channel are estimated and subtracted from the selected data events.

\subsection{\label{sec:resdecay} \boldmath{ Numbers of events from the $J/\psi$ and $\psi(3686)$ decays into  $p \bar p $}  }
\begin{figure}[h]
\includegraphics[height=7cm,width=9cm]{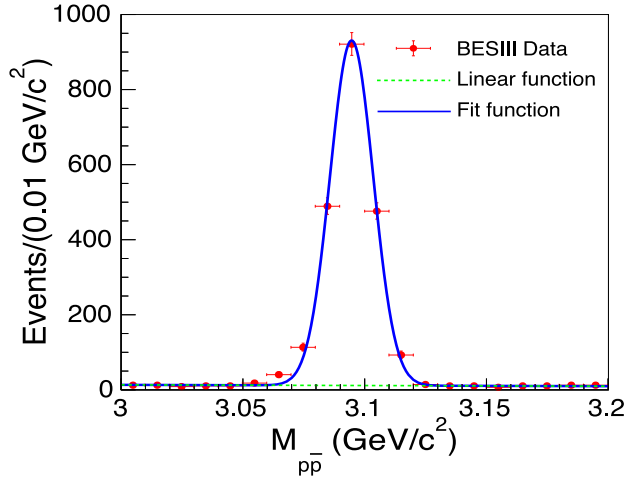}
\caption{Distribution of $M_{p\bar p }$ in the region of the $J/\psi$ resonance, for the data collected at $\sqrt{s}=3.773$~GeV.  The curves are the results of the fit. The  dashed green curve represents the linear fit function and the solid blue curve represents the sum of the Gaussian  (for resonance events) and the linear [for signal (Fig.~\ref{isrfeyn}) and  background events] functions. }
\label{invmassjpsi3773}
\end{figure}

\begin{figure}[h]
\includegraphics[scale=0.54]{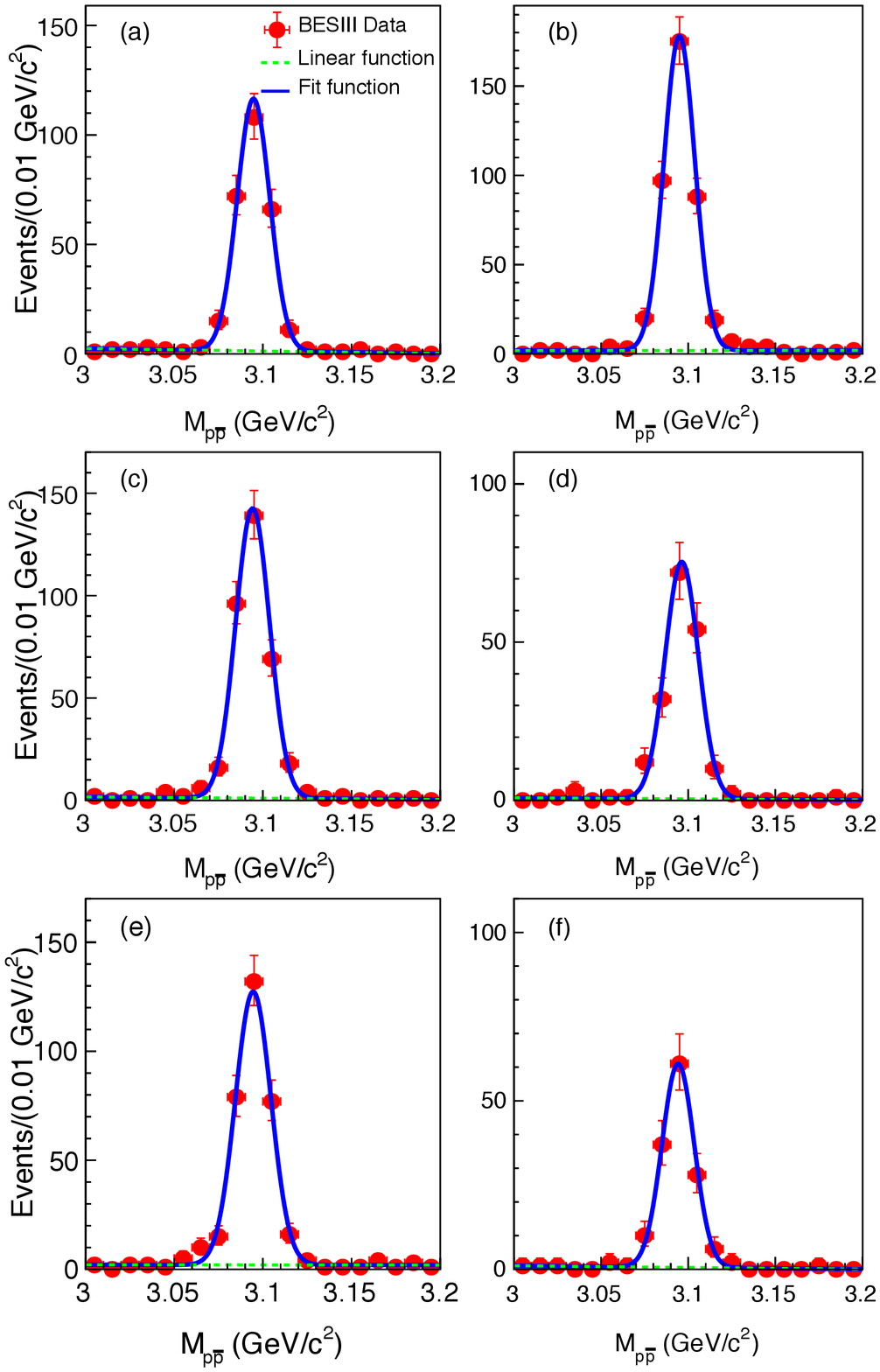}
\caption{Distributions of $M_{p\bar p }$  in the region of the $J/\psi$ resonance, for the data collected at $\sqrt{s}=$ (a) 4.008, (b) 4.226, (c) 4.258, (d) 4.358, (e) 4.416, and (f) $4.600$~GeV.  The curves are the result of the fits.   At  each c.m. energy, the numbers of resonance events and nonresonance events are determined. The  dashed green curve represents the linear fit function and the solid blue curve represents the sum of the Gaussian (for resonance events) and the linear [for signal (Fig.~\ref{isrfeyn}) and  background events] functions.}
\label{invmassjpsi}
\end{figure}

\begin{figure}[h]
\includegraphics[scale=0.54]{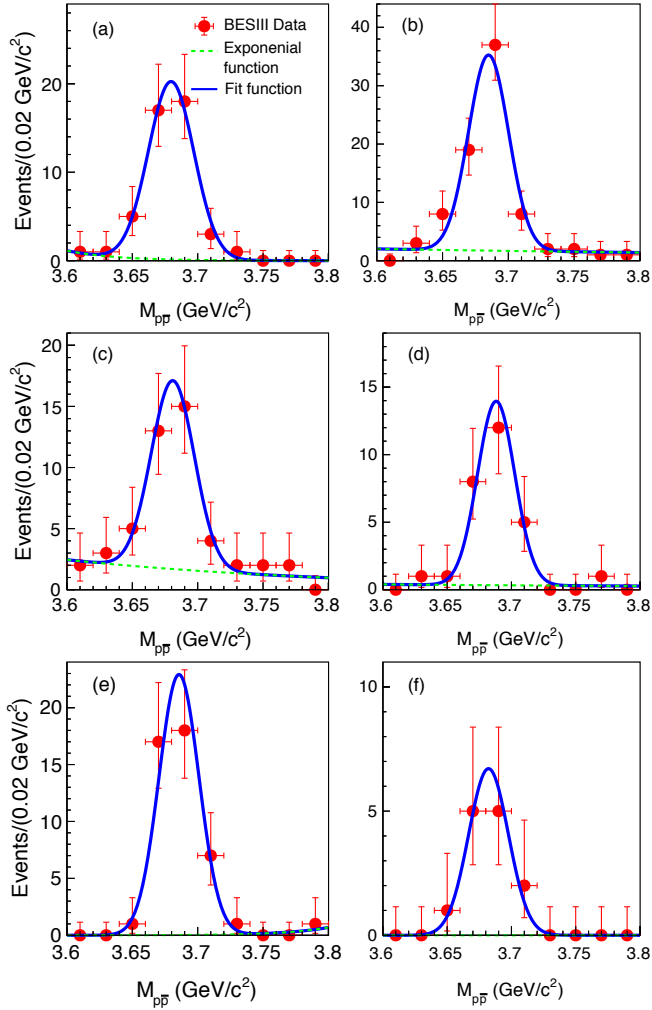}
\caption{Distributions of $M_{p\bar p }$ in the region of the $\psi(3686)$ resonance, for the data collected at $\sqrt{s}=$ (a) 4.008, (b) 4.226, (c) 4.258, (d) 4.358, (e) 4.416, and (f) $4.600$~GeV.  The curves are the results of the fits.   At  each c.m. energy, the numbers of resonance events and nonresonance events are determined.  The  dashed green curve represents the exponential fit function  and the solid blue curve represents the sum of the Gaussian  (for resonance events) and the exponential [for signal (Fig.~\ref{isrfeyn}) and  background events] functions.}
\label{invmasspsi2s}
\end{figure}

The selected events with $M_{p\bar p }$ falling in the regions of $J/\psi$ resonance are shown in Figs.~\ref{invmassjpsi3773} and \ref{invmassjpsi} and those in  $\psi(3686)$  resonance  in  Fig.~\ref{invmasspsi2s}.  The selected events for the different data samples  are fitted  using the sum of a Gaussian function (for resonance events) and a linear or exponential function (for signal  and possible remaining background channels).  The fit parameters are the number of resonance events, the number of nonresonance events, the constant of the linear/exponential function, the mean and the sigma of the Gaussian function.  The numbers of resonance and nonresonance events are calculated for each data sample separately. The numbers of  events for the  $J/\psi \to p \bar p $  and  $\psi(3686) \to p \bar p $ decays are listed in Table~\ref{tabresonnance}.

\begin{table}[h]
\begin{center}
\caption[]{Numbers of events for $J/\psi \to p \bar p$  and  $\psi(3686) \to p \bar p $ decays for the different data samples collected at the 7 c.m. energies. The analysis described in this paper  requires the emission of a hard ISR photon in the signal channel and is therefore not suitable to measure the number of events for the $\psi(3686) \to p \bar p $ decay at $\sqrt{s}=3.773$~GeV.}
\begin{ruledtabular}
\begin{tabular}{ccc}
$\sqrt{s}$ [GeV] &  $N_{J/\psi \to p \bar p}$ & $N_{\psi(3686) \to p \bar p }$ \\\hline
3.773 &  2046 $\pm$ 46 & $—---$ \\
4.008 &  266 $\pm$ 17 & 43.9 $\pm$ 7.3 \\
4.226 &  391 $\pm$ 20 & 64.1 $\pm$ 9.4  \\
4.258 &  340 $\pm$ 19 & 32.0 $\pm$ 7.3  \\
4.358 &  179 $\pm$ 14 & 24.7 $\pm$ 5.2  \\
4.416 &  317 $\pm$ 18 & 43.8 $\pm$ 6.6  \\
4.600 &  140 $\pm$ 12 & 13.0 $\pm$ 3.3  \\
\end{tabular}
\label{tabresonnance}
\end{ruledtabular}
\end{center}
\end{table}

\subsection{\label{sec:pionbkg} \boldmath {Background from $e^+ e^- \to p \bar p \pi^0 (\gamma)$}}
\begin{figure*}[th!]
\captionsetup[subfloat]{position=top,labelformat=empty}
  \subfloat[\label{fig:dsigma17}]{\resizebox{0.34\textwidth}{!}{ \includegraphics{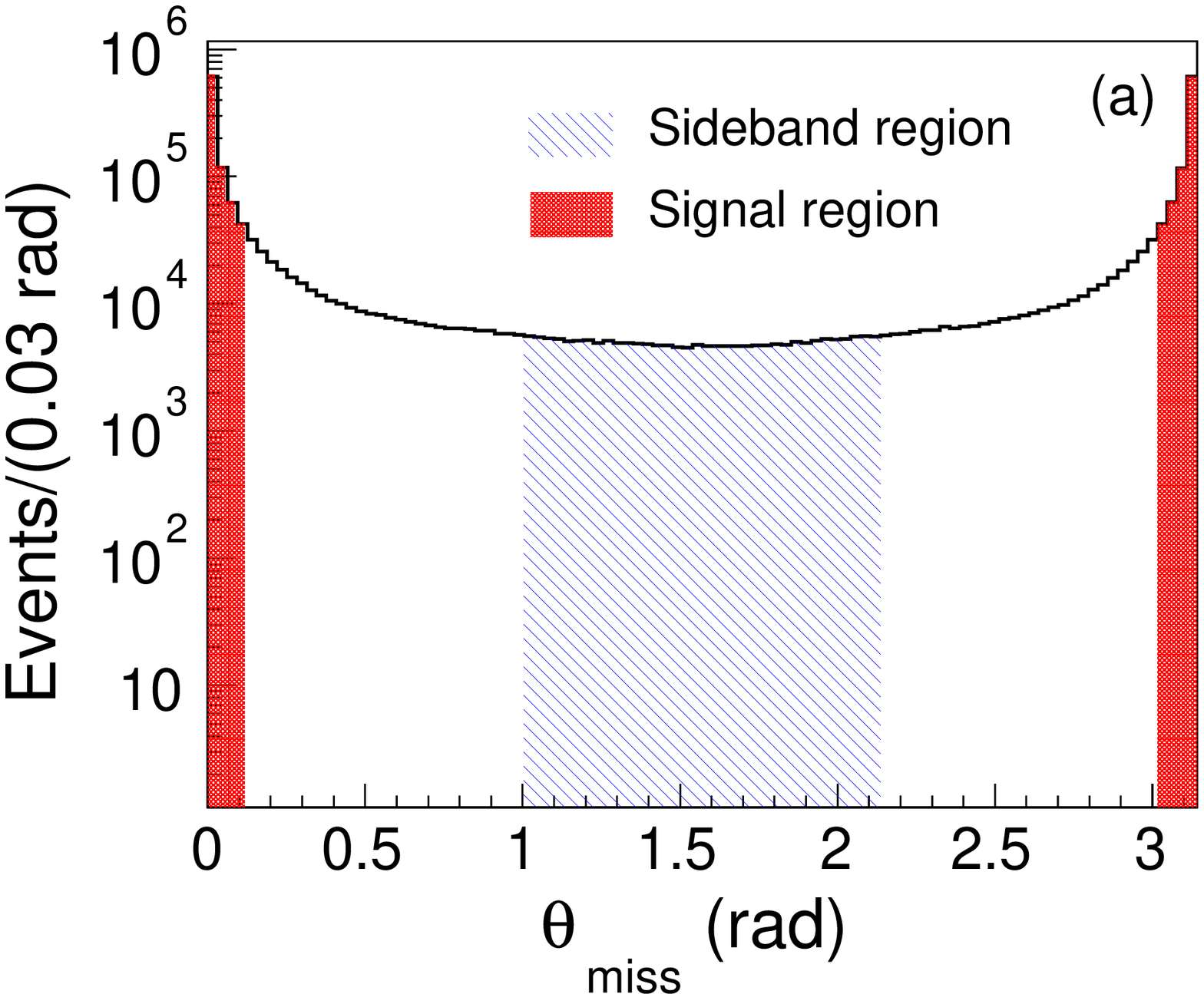} } }
  \subfloat[\label{fig:dsigma278}]{\resizebox{0.34\textwidth}{!}{ \includegraphics{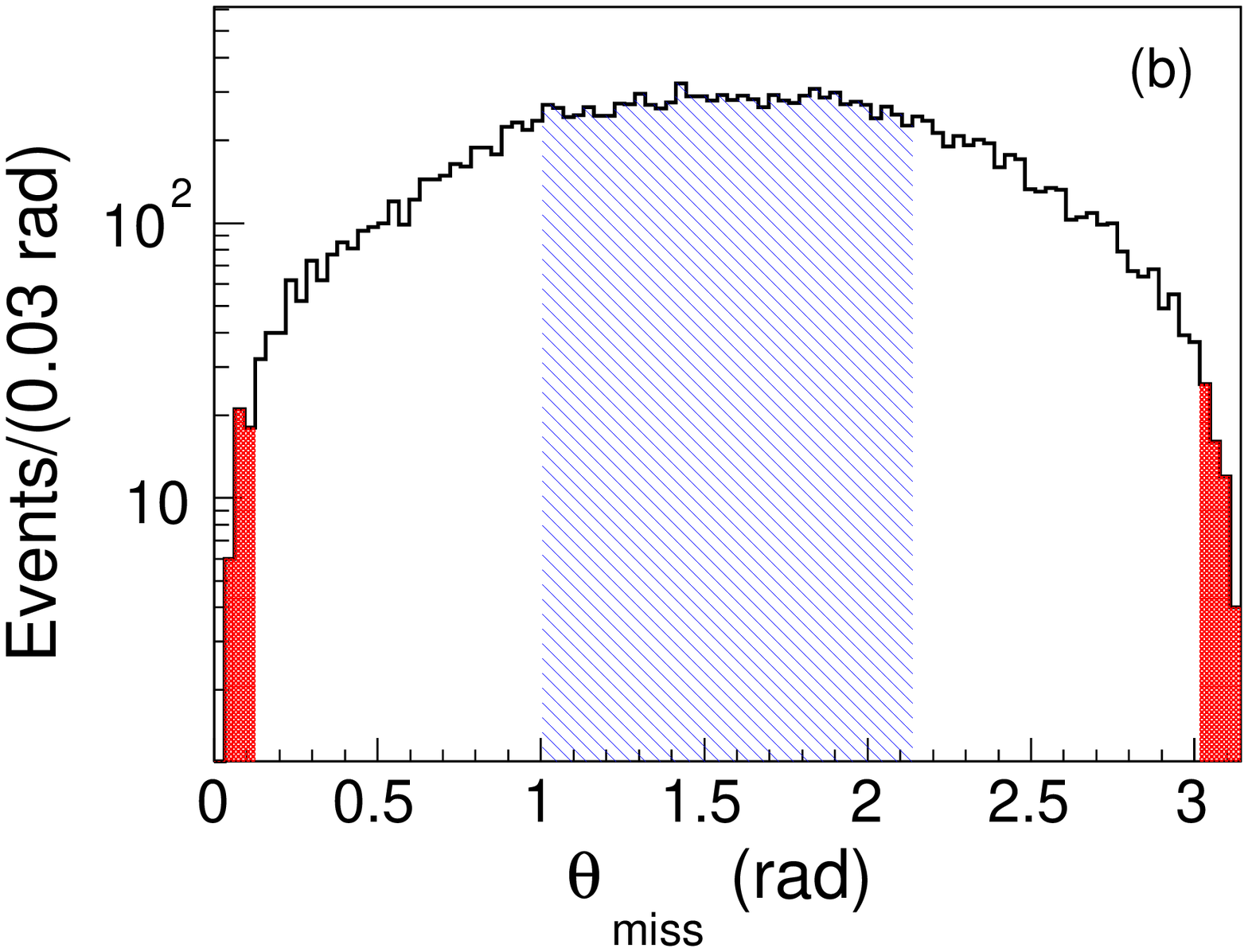} } }
 \subfloat[\label{fig:dsigma33}]{\resizebox{0.34\textwidth}{!}{ \includegraphics{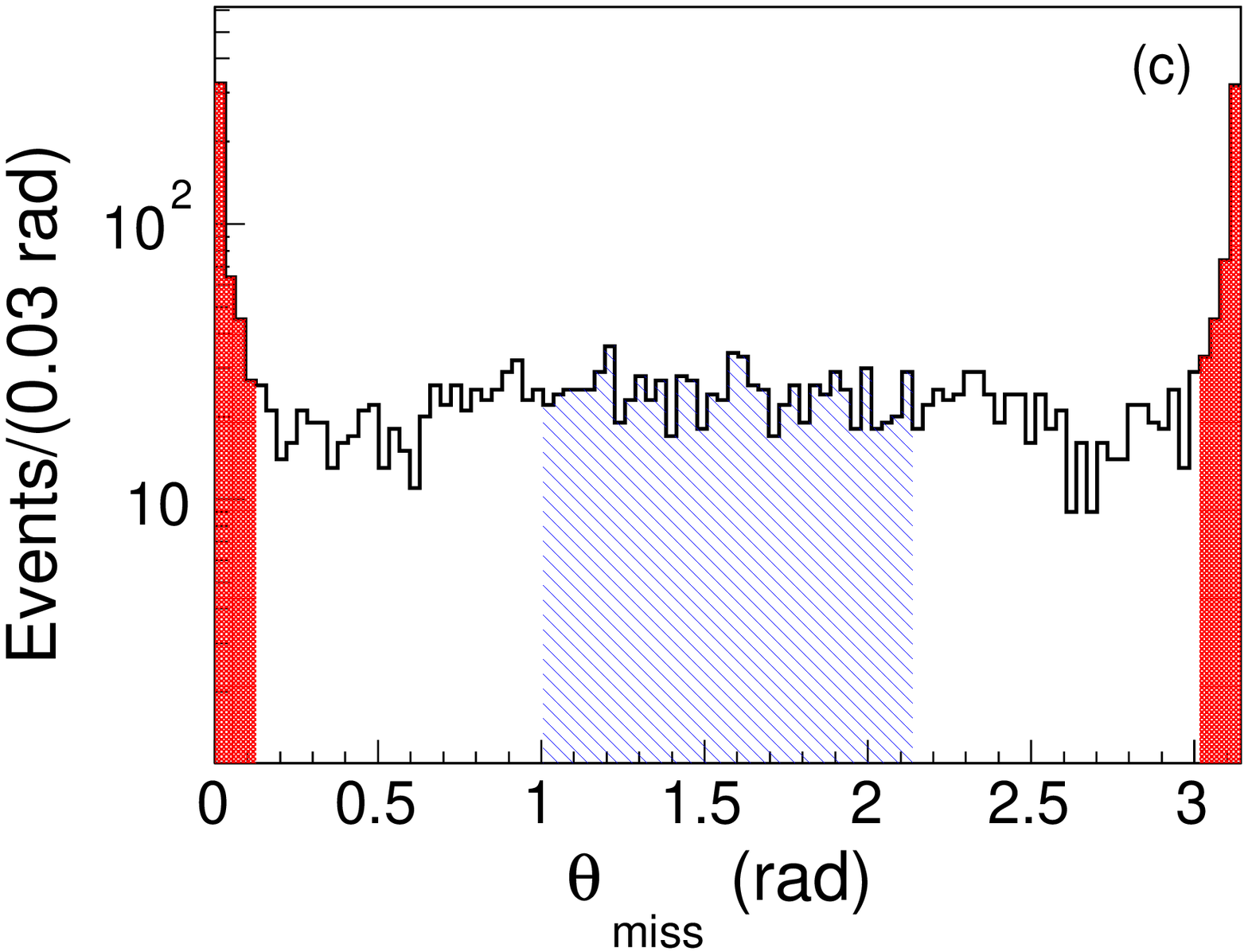} } } \\
 \caption{Distributions of  $\theta_{\rm miss}$  in the $M_{p\bar p }$ interval [2.0 - 3.0]~GeV/$c^2$, after applying the event selection conditions (except the condition on $\theta_{\rm miss}$) for the simulated  signal events (a), simulated background  $e^+ e^- \to p \bar p  \pi^0$  events (b), and  data events (c) at $\sqrt{s}=4.226$~GeV. The red and blue areas  represent the signal  and the sideband regions, respectively.}
  \label{fig:sideband}
  \end{figure*}

The process $e^+ e^- \to p \bar p \pi^0 (\gamma)$ is a critical background to the signal process since it contains the same detected charged particles, proton and antiproton, as the signal.  To estimate the background from the process $e^+ e^- \to p \bar p  \pi^0 (\gamma)$, we use  the difference of the $\theta_{\rm miss}$ distributions between signal and background events. The MC samples generated based on the measured angular distributions of the process $e^+ e^- \to p \bar p \pi^0$  \cite{Ablikim:2014kxa,Ablikim:2017gtb} are used.  Figure~\ref{fig:sideband} shows the distributions  of  $\theta_{\rm miss}$, the polar angle of the missing momentum, for data events and simulated signal and  $e^+ e^- \to p \bar p  \pi^0$ background events.  The red  (blue) area in Fig.~\ref{fig:sideband} represents the signal  (sideband) region.  The number of data events in the sideband region  ($N_2$)  and the number of background events in the signal region ($N_{\rm bkg}$)  are related by:
\be
N_{\rm bkg}=\frac{N_2-\beta_{\rm sig} N_1}{\beta_{\rm bkg}-\beta_{\rm sig}},
\label{eqsb1}
\ee
where $N_1$ is the number of data events in the signal region. The numbers $N_1$ and $N_2$ are determined from data after applying the event selection conditions except the $\theta_{\rm miss}$ requirement. The ratios $\beta_{\rm sig}$  and $\beta_{\rm bkg}$ are the $N_2/N_1$ ratios from the MC signal and background events, respectively. ISR effects ($e^+ e^- \to p \bar p \pi^0 \gamma$) are simulated with  the generator software package {\sc conexc} and they are used to correct $\beta_{\rm bkg}$. 
Data-MC difference in the calculation of the ratios $\beta_{\rm sig}$ and $\beta_{\rm bkg}$, or presence of other background events in the sideband or the signal region,  can provide wrong number of $N_{\rm bkg}$. These effects are considered in the calculation of the systematic uncertainty on the number
of selected $e^+ e^- \to p \bar p  \gamma$ events.

The number of background events $N_{p\bar{p}\pi^0 (\gamma)}$ is determined for each data sample separately.  This background source constitutes $2.3\%$ of the selected data events.

\subsection{\label{sec:twophoton} Background from two-photon channel}
\begin{figure*}[th!]
\captionsetup[subfloat]{position=top,labelformat=empty}
  \subfloat[\label{fig:dsigma17}]{\resizebox{0.34\textwidth}{!}{ \includegraphics{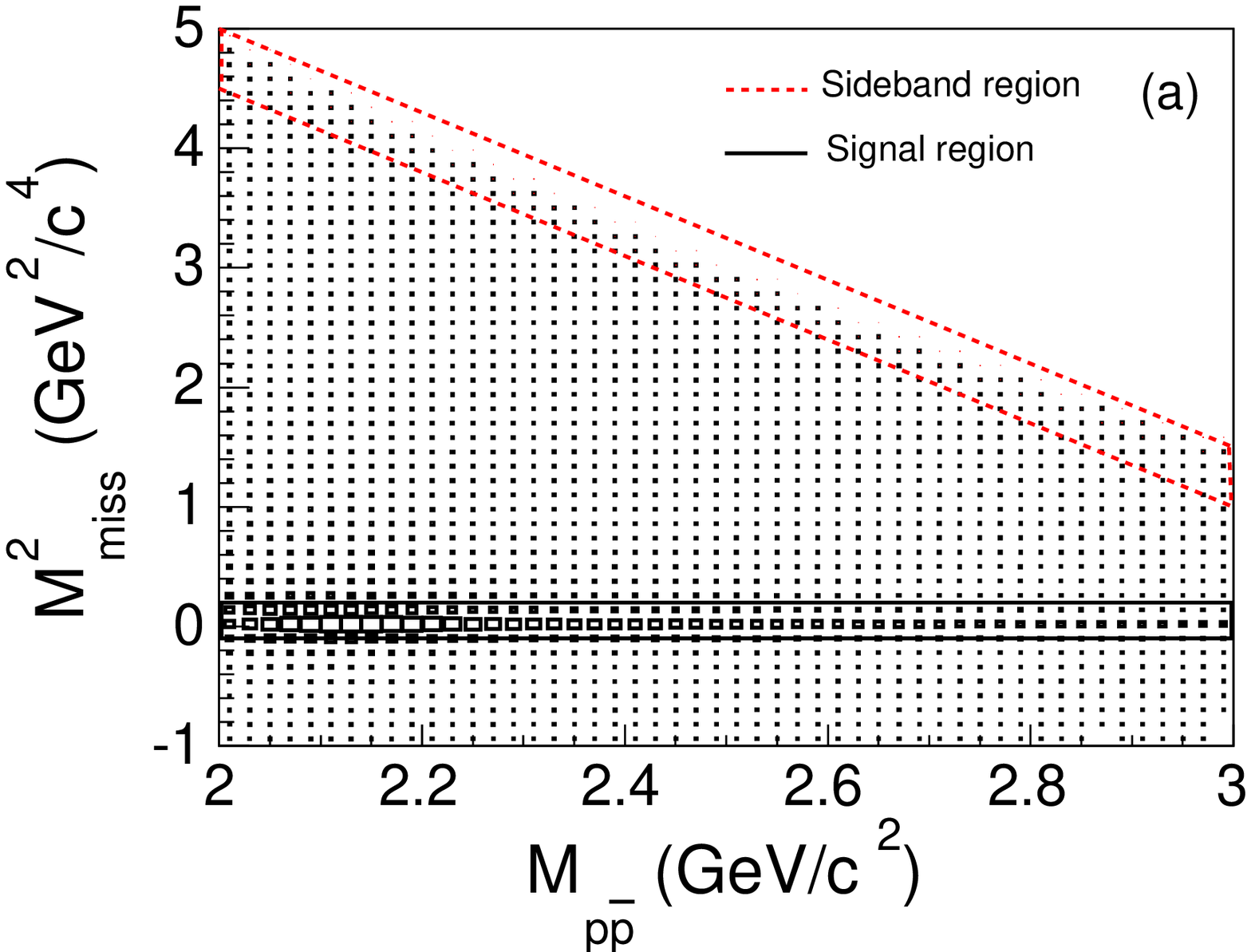} } }
  \subfloat[\label{fig:dsigma278}]{\resizebox{0.34\textwidth}{!}{ \includegraphics{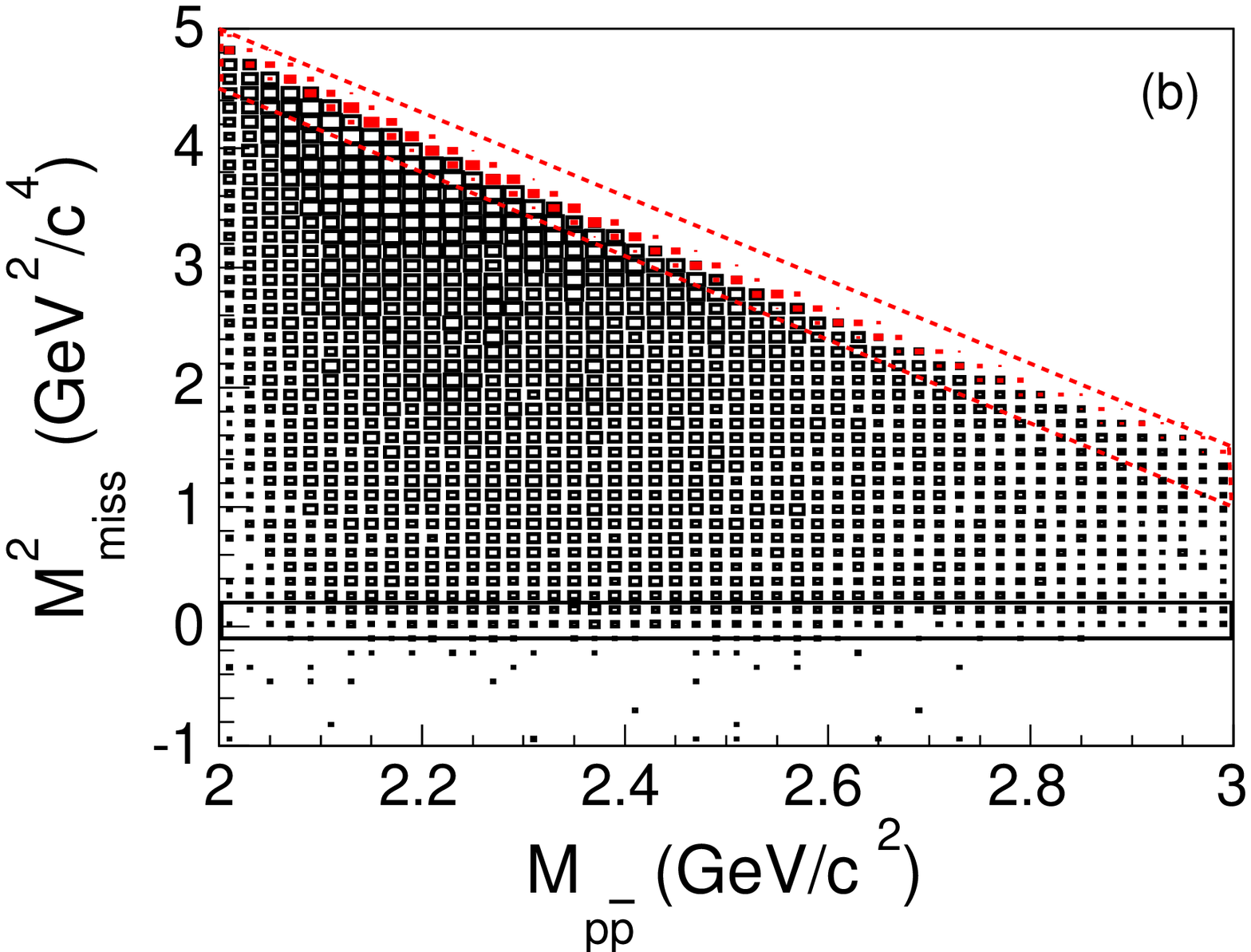} } }
  \subfloat[\label{fig:dsigma33}]{\resizebox{0.34\textwidth}{!}{ \includegraphics{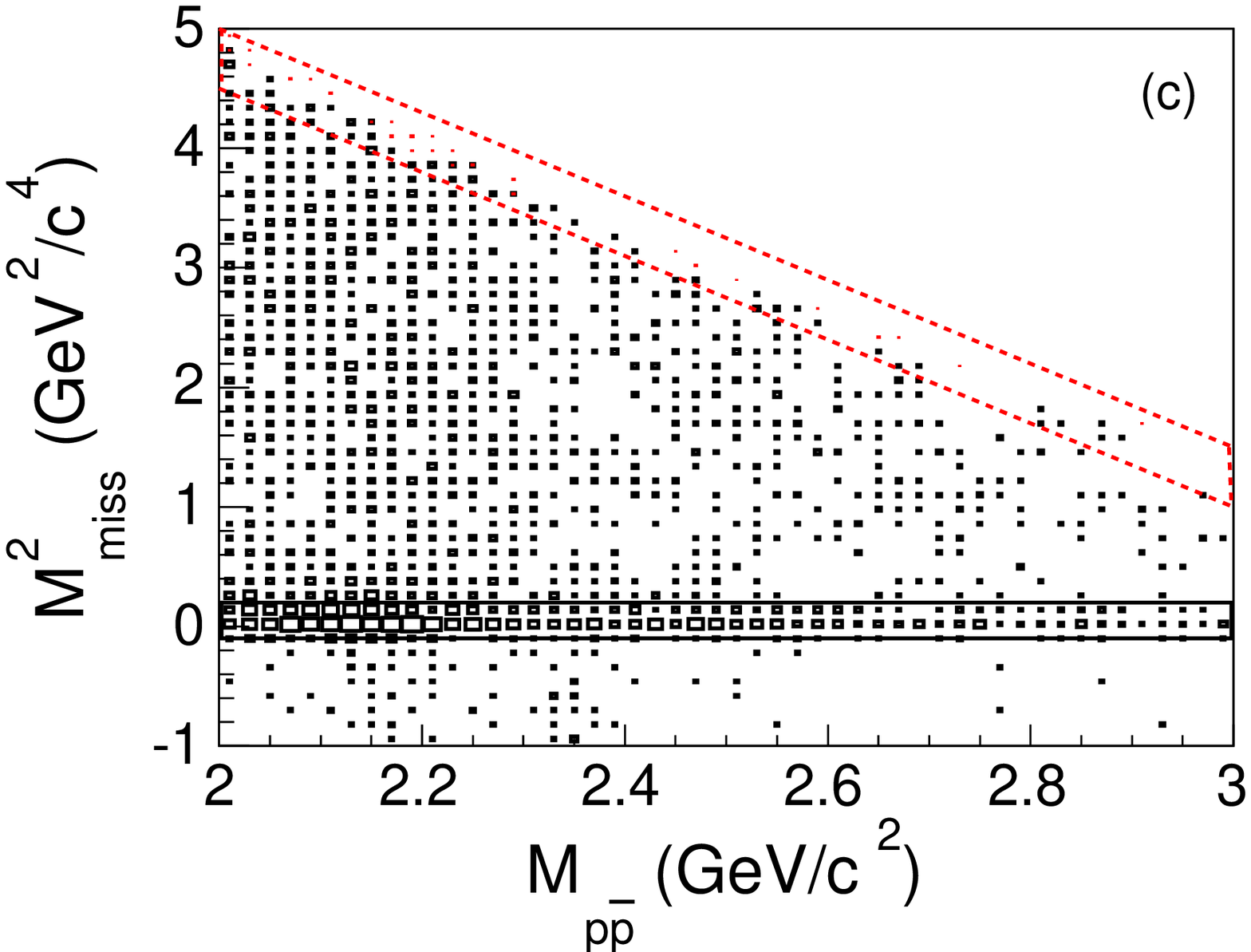} } }
\caption{Distributions of  $M_{\rm miss}^2$ versus   $M_{p\bar{p}}$, after applying the event selection conditions (except the condition on $M_{\rm miss}^2$) for the simulated  signal events (a),  two-photon  events (b), and  data events (c) at $\sqrt{s}=4.226$~GeV. The black solid lines represent the borders of the signal region.  The red filled squares describe the selected events of the sideband region ($|\vec{p}_{miss}|<0.25$~GeV/$c$).}
  \label{2sidbandtg}
 \end{figure*}

The number of background events from the two-photon channel $N_{2\gamma}$ is estimated using the same method described in Sec.~\ref{sec:pionbkg}.  Figure~\ref{2sidbandtg} shows the two-dimensional distributions of $M_{\rm miss}^2$ versus $M_{p\bar{p}}$ for the MC signal and two-photon events, and for the data events at $\sqrt{s}=4.226$~GeV.
 The region of large $M_{\rm miss}^2$ values ($|\vec{p}_{\rm miss}|<0.2$~GeV/$c$ at $\sqrt{s}>3.773$~GeV and $|\vec{p}_{\rm miss}|<0.25$~GeV/$c$ at $\sqrt{s}>4$~GeV) is chosen as the sideband region.
The black lines in Fig.~\ref{2sidbandtg} show the borders of the signal region at $\sqrt{s}$=4.226~GeV. The total number of background events from the two-photon channel constitutes $1.0\%$ of the total selected data events. No background events from the two-photon channel are survived in the $M_{p\bar p }$ region above 3.0~GeV/$c^2$.

The sum of the background events over the 7 c.m. energy points  for the  $e^+ e^- \to p \bar p \pi^0 (\gamma)$ and two-photon channels in each $M_{p\bar p}$ interval is given  in Table~\ref{tabbkgtot}.

\begin{table}[h]
\begin{center}
\caption[]{Differential luminosity ($L_{i}$),   numbers of background events  ($N_{\rm bkg}$) from $e^+ e^- \to p \bar p \pi^0$ and two-photon channel,  and   numbers of selected  events after background subtraction ($N_{\rm data}$) at each $M_{p\bar p }$ interval, from the combined data collected at the 7 c.m. energies. The numbers of events in the $M_{p\bar p }$ intervals [3.0 - 3.2]~GeV/$c^2$ and  [3.6 - 3.8]~GeV/$c^2$ are determined from the fits described in Sec.~\ref{sec:resdecay} and do  not include the background events from  the $J/\psi \to p \bar p $  and  $\psi(3686) \to p \bar p $ decays. The uncertainties are statistical.}
\begin{ruledtabular}
\begin{tabular}{ccccc}
$M_{p \bar p }$ [GeV/$c^2$] &  $L_{i}$ [pb$^{-1}$] & $N_{p\bar{p}\pi^0 (\gamma)}$ & $N_{2\gamma}$ & $N_{\rm data}$ \\\hline

2.000 - 2.025 & 2.39 &    5.0   $\pm$ 1.7 &   0.92 $\pm$ 0.80  &   218 $\pm$ 15 \\
2.025 - 2.050 & 2.59 &    4.2   $\pm$ 1.7 &   0.77 $\pm$ 0.45  &   343 $\pm$ 19 \\ 
2.050 - 2.075 & 2.65 &    7.2   $\pm$ 2.0 &   2.18 $\pm$ 0.87  &   380 $\pm$ 20 \\ 
2.075 - 2.100 & 2.72 &    4.6   $\pm$ 1.6 &   1.52 $\pm$ 0.77  &   467 $\pm$ 22 \\ 
2.100 - 2.125 & 2.79 &    4.6   $\pm$ 1.5 &   2.6 $\pm$ 1.1    &   456 $\pm$ 22 \\ 
2.125 - 2.150 & 2.86 &    5.2   $\pm$ 1.5 &   0.83 $\pm$ 0.57  &   491 $\pm$ 22 \\ 
2.150 - 2.175 & 2.93 &    7.8   $\pm$ 2.0 &   3.1 $\pm$ 1.2    &   455 $\pm$ 22 \\ 
2.175 - 2.200 & 3.00 &    6.0   $\pm$ 1.6 &   6.1 $\pm$ 2.1    &   409 $\pm$ 21 \\ 
2.200 - 2.225 & 3.08 &    8.9   $\pm$ 2.0 &   4.4 $\pm$ 1.4    &   338 $\pm$ 19 \\ 
2.225 - 2.250 & 3.16 &    5.6   $\pm$ 1.6 &   4.1 $\pm$ 1.6    &   300 $\pm$ 18 \\ 
2.250 - 2.275 & 3.24 &    4.9   $\pm$ 1.9 &   2.7 $\pm$ 1.2    &   227 $\pm$ 15 \\ 
2.275 - 2.300 & 3.32 &    7.5   $\pm$ 2.3 &   3.4 $\pm$ 1.3    &   199 $\pm$ 15 \\ 
2.300 - 2.350 & 6.91 &    9.0   $\pm$ 2.0 &   3.8 $\pm$ 1.4    &   303 $\pm$ 18 \\ 
2.350 - 2.400 & 7.28 &    16.7  $\pm$ 3.5 &   4.1 $\pm$ 1.8    &   279 $\pm$ 18 \\ 
2.400 - 2.450 & 7.69 &    6.1   $\pm$ 1.4 &   3.8 $\pm$ 1.5    &   322 $\pm$ 18 \\ 
2.450 - 2.500 & 8.13 &    5.5   $\pm$ 1.3 &   4.8 $\pm$ 2.1    &   281 $\pm$ 17 \\ 
2.500 - 2.550 & 8.60 &    5.4   $\pm$ 1.1 &   6.6 $\pm$ 2.2    &   204 $\pm$ 15 \\ 
2.550 - 2.600 & 9.12 &    2.68   $\pm$ 0.70 &  5.7 $\pm$ 2.1    &   193 $\pm$ 14 \\ 
2.600 - 2.650 & 9.68 &    5.6   $\pm$ 1.5 &   3.3 $\pm$ 1.6    &   146 $\pm$ 13 \\ 
2.650 - 2.700 & 10.30 &    3.7  $\pm$ 1.0 &   2.3 $\pm$ 1.3    &   123 $\pm$ 11 \\ 
2.700 - 2.750 & 10.97 &    4.5  $\pm$ 1.4 &   1.4 $\pm$ 1.1    &   121 $\pm$ 11 \\ 
2.750 - 2.800 & 11.72 &    6.0  $\pm$ 1.6 &   0.00 $\pm$ 0.10  &   115 $\pm$ 11 \\ 
2.800 - 2.850 & 12.54 &    4.5  $\pm$ 1.3 &   0.46 $\pm$ 0.64  &   98 $\pm$ 10 \\ 
2.850 - 2.900 & 13.46 &    6.0  $\pm$ 1.8 &   1.3 $\pm$ 1.2    &   100 $\pm$ 11 \\ 
2.900 - 2.950 & 6.44 &     2.03 $\pm$ 0.43 &   2.2 $\pm$ 1.5   &   36.8 $\pm$ 6.6 \\ 
2.950 - 3.000 & 6.84 &     1.05 $\pm$ 0.38 &   0 $\pm$ 0   &   40.0 $\pm$ 6.4 \\
3.000 - 3.200 & 32.23 &    3.54 $\pm$ 0.61 &    0 $\pm$ 0   &   145 $\pm$ 15 \\
3.200 - 3.400 & 42.91 &    4.10 $\pm$ 0.63 &    0 $\pm$ 0   &   66.9 $\pm$ 8.4 \\
3.400 - 3.600 & 60.36 &    2.51 $\pm$ 0.45 &    0 $\pm$ 0   &   52.5 $\pm$ 7.4 \\
3.600 - 3.800 & 87.18 &    3.24 $\pm$ 0.47 &    0 $\pm$ 0   &   41 $\pm$ 12 \\
\end{tabular}
\label{tabbkgtot}
\end{ruledtabular}
\end{center}
\end{table}

\section{\label{sec:eff} Signal efficiency}

\begin{figure}[h]
\includegraphics[height=7cm,width=9cm]{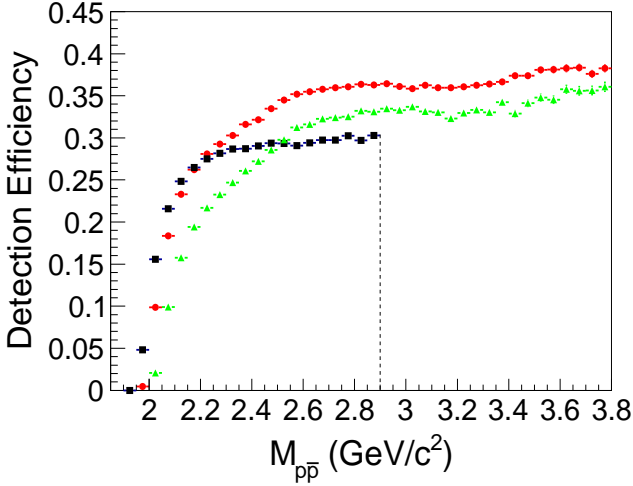}
\caption{Efficiency of the signal $e^+ e^- \to p \bar p  \gamma$  as a function of $M_{p\bar p }$  for $\sqrt{s}=3.773$~GeV (black squares),  $\sqrt{s}=4.226$~GeV (red points), and  $\sqrt{s}=4.600$~GeV (green triangles).}
\label{Eff4230}
\end{figure}

The signal efficiency is determined from the MC simulations of the signal by dividing the number of selected events by the number of generated events. The signal events are generated in the full range of the proton momenta and the photon polar angle. The integrated signal efficiency at $\sqrt{s}=3.773$~GeV is equal to $16.8\%$. It decreases to $12.6\% $ at the highest c.m. energy ($\sqrt{s}=4.600$~GeV). The signal efficiency is determined in each $M_{p\bar p }$ interval using the MC events of the  process $e^+ e^- \to p \bar p  \gamma$  generated up to the next-to-leading order radiative corrections. The parametrizations for $G_E$ and $G_M$ from Ref.~\cite{Czyz:2014sha} are used to calculate the efficiency of the signal. The $M_{p\bar p }$  dependence of the signal efficiency is shown in Fig.~\ref{Eff4230} for $\sqrt{s}=3.773$, $4.226$, and $4.600$~GeV.  In the low $M_{p\bar p }$  region ($M_{p \bar p }<2$~GeV/$c^2$), the proton and antiproton are produced in a narrow cone around the vector opposite to the direction of the ISR photon.  The signal events at low  $M_{p\bar p }$ region are suppressed due to the limited acceptance of the BESIII tracking system. 
 

\section{\label{sec:totalxs} Cross section for the process \boldmath{\lowercase{ $e^+ e^- \to p \bar p $}}  and the proton effective FF}
The Born cross section for the process $e^+ e^- \to p \bar p  $  is calculated in each $M_{p\bar p }$ interval $i$ and for each data sample $j$  ($j=1,2,...,7$) as follows:

\be
\sigma_{ij}=\frac{N_{ij}}{\epsilon_{ij} (1+\delta_{ij}) L_{ij} },
\ee
where  $N_{ij}$ is the number of  selected $e^+ e^- \to p \bar p  \gamma$ events after background subtraction, $\epsilon_{ij}$ is the detection efficiency,  $(1+\delta_{ij})$ is the radiative correction factor and $L_{ij}$ is the ISR differential luminosity.  The  index $j$ runs over the 7 c.m.  energies.

The detection efficiency $\epsilon_{ij}$ is determined in each $M_{p\bar p }$  interval using the MC events of the  process $e^+ e^- \to p \bar p  \gamma$  generated up to the next-to-leading order radiative corrections.  The radiative correction factor $(1+\delta_{ij})$ describes the distortion of the $e^+ e^- \to p \bar p  \gamma$ cross section due to contribution of  higher order diagrams. It is calculated using the generated MC events of the signal and takes into account vacuum polarization and photon emissions from the initial and final states.    The differential luminosity  $L_i$ is  calculated as:
\be
L_{ij}=\int W(s_j, x_{ij}) {\cal {L}}_j dx_{ij},~ x_{ij}=1-\frac{q^2_{ij}}{s_j},
\label{diffli}
\ee
where $W(s_j,x_{ij})$ [Eq.~(\ref{eq:isrc2})] is a function of the c.m. energy squared $s_j$ ($j=1,2,...,7$) and the energy fraction $x_{ij}$, and  ${\cal {L}}_j$  is the integrated luminosity collected at the c.m. energy $\sqrt{s_j}$ (Table~\ref{dataset}).  The integral in Eq.~(\ref{diffli}) is performed over the width of the selected $M_{p\bar p }$ interval. The MC events of the signal process are used to determine the  $p\bar p$ mass resolution in each $M_{p\bar p }$ interval.  The width of the chosen  $M_{p\bar p }$ interval exceeds the mass resolution for all the $p\bar p$ masses.

The Born cross sections $\sigma_{ij}$  are combined using the error weighted combination method \cite{1402}:

\begin{equation} \label{eqxx}
\begin{split}
\sigma _{p \bar p }(M_{p\bar p })&=\sigma_i=   \Sigma_j (w_{ij} \sigma_{ij}),~\Delta \sigma_i=\sqrt{\frac{1}{\Sigma_j W_{ij}}}, \\
w_{ij}&=\frac{W_{ij}}{\Sigma_l W_{il}},~W_{il}=\frac{1}{(\Delta \sigma_{il})^2},
\end{split}
\end{equation}
where $\Delta \sigma_{i}$ and $\Delta \sigma_{ij}$ are the statistical errors of  $\sigma_{i}$ and $\sigma_{ij}$, respectively. The  indices $j$ and $l$ run over the 7 c.m.  energies.

The obtained values of the Born cross section for the process  $e^+ e^- \to p \bar p $  are listed in Table~\ref{tabResEffESSI2}. The quoted uncertainties are statistical and systematic.  The systematic uncertainties of  the measured  cross section include uncertainties from tracking, PID, $E_{\rm EMC}/p_{\rm rec}$  requirement, background estimation, $M_{\rm miss}^2$ and $\theta_{\rm miss}$ requirements, and luminosity determination. The contributions of the uncertainties from the tracking of the two charged particles ($2.0\%$), PID ($2.0\%$) and $E_{\rm EMC}/p_{\rm rec}$  requirement ($1.0\%$) are uniform over the considered $M_{p\bar p }$  range \cite{Ablikim:2015vga}. To determine the uncertainty from the background estimation of the  $e^+ e^- \to p \bar p  \pi^0$ and two-photon channels, we calculate the number of selected events (before efficiency correction) with and without background subtraction. The difference between the two cases (1.0$\%$-7.3$\%$ for the $e^+ e^- \to p \bar p  \pi^0$ channel  and less than 5.4$\%$  for the two-photon channel)  is taken as systematic uncertainty  from the background estimation. We associate 0.5$\%$ systematic uncertainty to the possible background contribution from $e^+e^-\to\gamma {\cal{R}}~({\cal{R}} \to p \bar p  \gamma),~{\cal{R}}= J/\psi, \psi(3686)$.
To study the systematic uncertainties from the $\theta_{\rm miss}$  and $M_{\rm miss}^2$ requirements, the Born cross section for the process $e^+ e^- \to p \bar p $  is recalculated using reduced selection windows of about $20\%$ compared to the original values. The uncertainties from the $\theta_{\rm miss}$  ($M_{\rm miss}^2$) requirements are found to be  less than  6$\%$ (5$\%$). The main sources of the systematic uncertainties on the measurements of the integrated luminosity at different c. m. energies are correlated \cite{Ablikim:2015orh, Ablikim:2015nan}. A conservative number of $0.8\%$ is taken as systematic uncertainty from the integrated luminosity measurements.  In addition, we associate 0.5$\%$ systematic uncertainty to the radiator function $W(s,x)$ \cite{Druzhinin:2011qd}  and 1.0$\%$ to the calculation of the  final state radiation  \cite{Czyz:2014sha}.  At low $M_{p\bar p }$  region, the uncertainty of the Born cross section is dominated by the uncertainty in the measured FF ratio ${\rm R}=|G_E|/|G_M|$. The values of the signal efficiency depend on the model of the proton FFs used in the event generator.  The model error due to the uncertainty in the measured ${\rm R}$ is determined by varying ${\rm R}$ within its statistical uncertainty (see Sec.~\ref{sec:FFratio}). It decreases from  $8\%$ at 2~GeV/$c^2$ to 3$\%$-4$\%$ in the   $M_{p\bar p }$  region below 3.0~GeV/$c^2$.  For $M_{p\bar{p}}>3$~GeV/$c^{2}$, where  ${\rm R}$ is not measured,  the model uncertainty  ($\sim 9\%$) is estimated as the difference between the detection efficiencies obtained with $|G_E|=0$ and $|G_M|=0$, divided by two. In each $M_{p\bar p}$ interval, the systematic uncertainties  listed above are added in quadrature.

\begin{figure}[h]
\includegraphics[height=7.3cm,width=9.5cm]{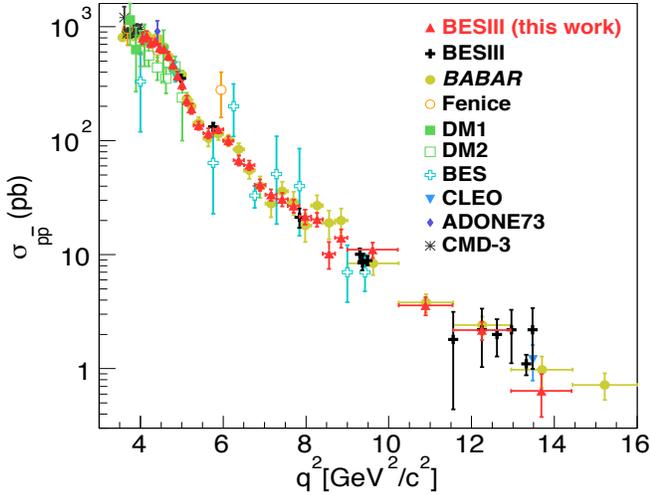}
\caption{Born cross section values for the process $e^+ e^-\to p \bar p $  measured in this analysis and in other $e^+ e^-$ experiments: Fenice~\protect\cite{Antonelli:1998fv}, DM1~\protect\cite{Delcourt:1979ed}, DM2~\protect\cite{Bisello:1983at,Bisello:1990rf}, BES~\protect\cite{Ablikim:2005nn}, BESIII~\protect\cite{Ablikim:2015vga}, CLEO~\protect\cite{Pedlar:2005sj}, {\it BABAR}~\protect\cite{Lees:2013rqd,Lees:2013ebn}, CMD-3~\protect\cite{Akhmetshin:2015ifg}, and ADONE73~\protect\cite{Castellano:1973}.}
\label{xstotpid}
\end{figure}

\begin{figure}[h]
\includegraphics[height=7.3cm,width=9.5cm]{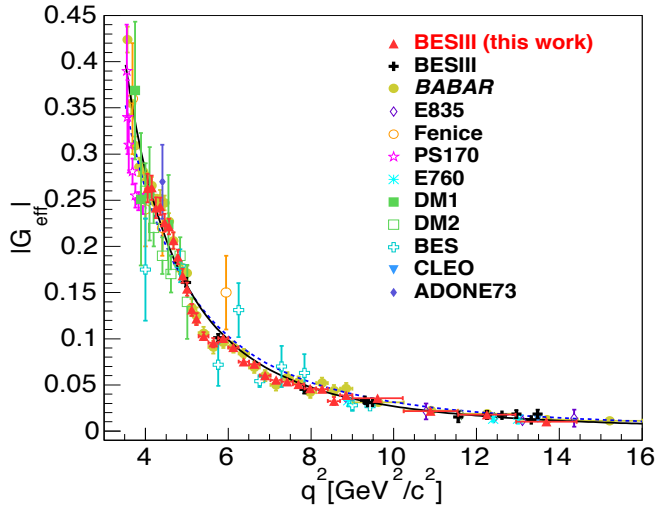}
\caption{Proton effective FF values measured  in this analysis and in other experiments: E835~\protect\cite{Andreotti:2003bt,Ambrogiani:1999bh}, Fenice~\protect\cite{Antonelli:1998fv}, PS170~\protect\cite{Bardin:1994am}, E760~\protect\cite{Armstrong:1992wq}, DM1~\protect\cite{Delcourt:1979ed}, DM2~\protect\cite{Bisello:1983at,Bisello:1990rf}, BES~\protect\cite{Ablikim:2005nn}, BESIII~\protect\cite{Ablikim:2015vga}, CLEO~\protect\cite{Pedlar:2005sj}, {\it BABAR}~\protect\cite{Lees:2013rqd,Lees:2013ebn},  and ADONE73~\protect\cite{Castellano:1973}. The blue dashed curve shows the QCD inspired parametrization~\cite{Shirkov:1997wi,Bianconi:2015vva} based on Eq.~(\ref{eqEffqcd}). The solid black curve shows the parametrization [Eq.~(\ref{rekalo})] suggested in Ref.~\cite{TomasiGustafsson:2001za}.}
\label{GEFFTOTPID}
\end{figure}

Knowing the Born cross section  for the process $e^+ e^- \to p \bar p $, one can determine the effective FF of the proton by
\be
|G_{\rm eff}|^2=\frac{2\tau|G_M|^2+|G_E|^2}{2\tau+1}=\frac{3 q^2 \sigma_{p \bar p}}{4 \pi \alpha^2 C (1+\frac{2 M_p^2}{q^2})}.
\label{eqEffdata}
\ee
The obtained values of $|G_{\rm eff}|$ are reported in Table~\ref{tabResEffESSI2}  for each $M_{p\bar p }$ interval. The results on the Born cross section and  the proton effective FF are shown in Figs.~\ref{xstotpid} and \ref{GEFFTOTPID}, respectively.  The results are consistent with previous experiments.
In particular, we reproduce the structures seen in the  measurements of the proton effective FF by the {\it BABAR} Collaboration ~\cite{Lees:2013rqd,Lees:2013ebn}. References~\cite{Ambrogiani:1999bh,Brodsky:2007hb,TomasiGustafsson:2001za,Shirkov:1997wi} provide several parametrizations of the timelike proton FFs. For example, the blue dashed curve in Fig.~\ref{GEFFTOTPID} represents the quantum chromodynamics (QCD) inspired parametrization of $|G_{\rm eff}|$ from Refs.~\cite{Shirkov:1997wi,Bianconi:2015vva}:
\be
|G_{\rm eff}|=\frac{{\cal{A_{\rm QCD}}}}{q^4 [\log^2(q^2/\Lambda_{\rm QCD}^2)+\pi^2]},
\label{eqEffqcd}
\ee
where the parameters ${\cal{A_{\rm QCD}}}=72~(\mbox{GeV/$c$})^4$ and $\Lambda_{\rm QCD}=0.52~(\mbox{GeV/$c$})$ are obtained from a fit to the previous experimental data \cite{Bianconi:2015owa}. The data on the timelike effective FF are best reproduced by the function proposed in Ref.~\cite{TomasiGustafsson:2001za},

\be
|G_{\rm eff}|=\frac{{\cal{A}}}{(1+q^2/m_a^2)[1-q^2/q_0^2 ]^2},~q_0^2=0.71~(\mbox{GeV/$c$})^2,
\label{rekalo}
\ee
where ${\cal{A}}=7.7$ and $m_a^2=14.8~(\mbox{GeV/$c$})^2$ are the fit parameters obtained previously in Ref.~\cite{Bianconi:2015owa}. It is illustrated in Fig.~\ref{GEFFTOTPID} by the solid black curve.

\begin{figure}[h]
\includegraphics[scale=0.45]{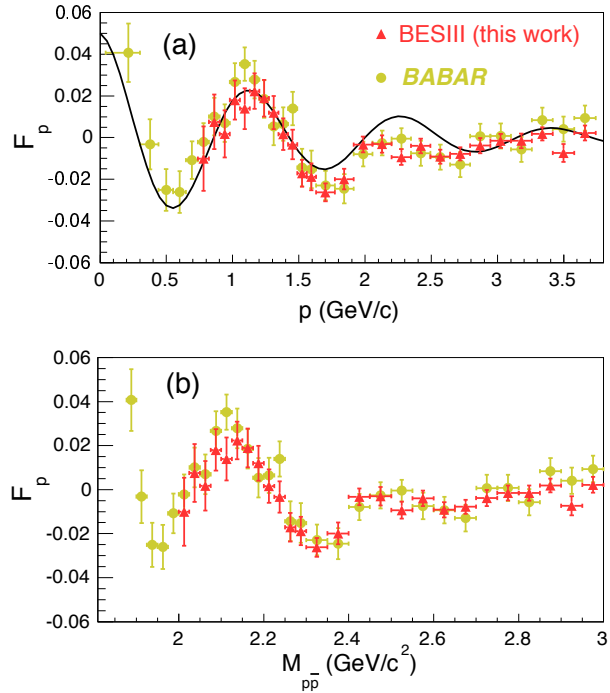}
\caption{Proton effective FF values, after subtraction of the smooth function described by Eq.~(\ref{rekalo}), as a function of the relative momentum ${\rm P}$ (a) and $M_{p\bar p }$  (b). The  data are from the present analysis and from {\it BABAR} experiment \protect\cite{Lees:2013ebn} measured in the $M_{p\bar p }$ intervals below 3~GeV/$c^2$. The black curve shows the parametrization from Ref.~\cite{Bianconi:2015owa} based on Eq.~(\ref{fosci}).}
\label{GEFFTOTPIDOSC}
\end{figure}

The two functions [Eqs.~(\ref{eqEffqcd}] and (\ref{rekalo})) reproduce the behavior of the effective FF over the long $q^2$ range. However, the measurements indicate some oscillating structures and therefore a more complex behavior   than the smooth decrease  predicted by QCD  as a function of $q^2$.  These oscillations are clearly seen when the data are plotted as a function of the 3-momentum ${p}$ of the relative motion of the final proton and antiproton \cite{Bianconi:2015vva}. Figure~\ref{GEFFTOTPIDOSC}(a) shows the values of the proton effective FF as a function of ${p}$  after subtraction of the smooth function described by Eq.~(\ref{rekalo}). The black solid curve in Fig.~\ref{GEFFTOTPIDOSC}(a) describes the periodic oscillations and has the form \cite{Bianconi:2015vva}
\be
F_{p}=A^{\rm osc} \exp(-B^{\rm osc} {p}) \cos (C^{\rm osc} {p} +D^{\rm osc}),
\label{fosci}
\ee
where $A^{\rm osc}=0.05$, $B^{\rm osc}=0.7~(\mbox{GeV/$c$})^{-1}$, $C^{\rm osc}=5.5~(\mbox{GeV/$c$})^{-1}$ and $D^{\rm osc}=0.0$ are obtained previously from a fit to the {\it BABAR} data \cite{Bianconi:2015owa}. The origin of these oscillating structures can be attributed to an interference effect involving rescattering processes in the final state \cite{Bianconi:2015vva} or to independent resonant structures, as in Ref.~\cite{Lorenz:2015pba}. The structure seen around $M_{p\bar p }=$ 2.15~GeV/$c^2$ [Fig.~\ref{GEFFTOTPIDOSC}(b)] can be for example attributed to the $\rho(2150)$ resonance \cite{Patrignani:2016xqp}. Other possible interpretations of these structures are not excluded here.

 \begin{table}[h]
\begin{center}
\caption[]{Born cross section of the  process $e^+ e^-\to p \bar p $  and the effective FF measured in each $M_{p\bar p }$ interval. The first and second uncertainties are statistical and systematic, respectively.}
\begin{ruledtabular}
\begin{tabular}{ccc}
$M_{p \bar p }$ [GeV/$c^2$] &     $\sigma _{p \bar p }~$[pb]&     $|G_{\rm eff}|$ \\\hline
 2.000 - 2.025 & 797 $\pm$ 56 $\pm$ 75 & 0.263 $\pm$ 0.009 $\pm$ 0.012 \\
 2.025 - 2.050 & 833 $\pm$ 46 $\pm$ 69 & 0.264 $\pm$ 0.007 $\pm$ 0.011 \\
 2.050 - 2.075 & 723 $\pm$ 38 $\pm$ 56 & 0.242 $\pm$ 0.006 $\pm$ 0.009 \\
 2.075 - 2.100 & 749 $\pm$ 35 $\pm$ 46 & 0.243 $\pm$ 0.006 $\pm$ 0.007 \\
 2.100 - 2.125 & 654 $\pm$ 31 $\pm$ 47 & 0.226 $\pm$ 0.005 $\pm$ 0.008 \\
 2.125 - 2.150 & 637 $\pm$ 29 $\pm$ 40 & 0.221 $\pm$ 0.005 $\pm$ 0.007 \\
 2.150 - 2.175 & 557 $\pm$ 27 $\pm$ 39 & 0.206 $\pm$ 0.005 $\pm$ 0.007 \\
 2.175 - 2.200 & 467 $\pm$ 24 $\pm$ 31 & 0.189 $\pm$ 0.005 $\pm$ 0.006 \\
 2.200 - 2.225 & 371 $\pm$ 21 $\pm$ 27 & 0.168 $\pm$ 0.005 $\pm$ 0.006 \\
 2.225 - 2.250 & 310 $\pm$ 19 $\pm$ 22 & 0.154 $\pm$ 0.005 $\pm$ 0.005 \\
 2.250 - 2.275 & 225 $\pm$ 16 $\pm$ 16 & 0.131 $\pm$ 0.005 $\pm$ 0.005 \\
 2.275 - 2.300 & 192 $\pm$ 14 $\pm$ 14 & 0.121 $\pm$ 0.005 $\pm$ 0.005 \\
 2.300 - 2.350 & 136.1 $\pm$ 8.1 $\pm$ 7.9 & 0.103 $\pm$ 0.003 $\pm$ 0.003 \\
 2.350 - 2.400 & 116.3 $\pm$ 7.5 $\pm$ 9.5 & 0.096 $\pm$ 0.003 $\pm$ 0.004 \\ 
 2.400 - 2.450 & 126.1 $\pm$ 7.2 $\pm$ 6.3 & 0.101 $\pm$ 0.003 $\pm$ 0.003 \\ 
 2.450 - 2.500 & 100.1 $\pm$ 6.2 $\pm$ 6.7 & 0.091 $\pm$ 0.003 $\pm$ 0.003 \\ 
 2.500 - 2.550 & 67.4 $\pm$ 5.0 $\pm$ 4.7 & 0.075 $\pm$ 0.003 $\pm$ 0.003 \\ 
 2.550 - 2.600 & 61.1 $\pm$ 4.6 $\pm$ 3.7 & 0.072 $\pm$ 0.003 $\pm$ 0.002 \\ 
 2.600 - 2.650 & 41.0 $\pm$ 3.7 $\pm$ 2.9 & 0.060 $\pm$ 0.003 $\pm$ 0.002 \\ 
 2.650 - 2.700 & 33.6 $\pm$ 3.2 $\pm$ 2.3 & 0.055 $\pm$ 0.003 $\pm$ 0.002 \\ 
 2.700 - 2.750 & 30.7 $\pm$ 3.0 $\pm$ 3.0 & 0.053 $\pm$ 0.003 $\pm$ 0.003 \\ 
 2.750 - 2.800 & 26.8 $\pm$ 2.7 $\pm$ 2.4 & 0.051 $\pm$ 0.003 $\pm$ 0.002 \\ 
 2.800 - 2.850 & 21.6 $\pm$ 2.3 $\pm$ 2.3 & 0.046 $\pm$ 0.002 $\pm$ 0.002 \\ 
 2.850 - 2.900 & 20.4 $\pm$ 2.2 $\pm$ 1.8 & 0.045 $\pm$ 0.002 $\pm$ 0.002 \\ 
 2.900 - 2.950 & 10.2 $\pm$ 2.2 $\pm$ 1.6 & 0.033 $\pm$ 0.004 $\pm$ 0.002 \\ 
 2.950 - 3.000 & 14.1 $\pm$ 2.4 $\pm$ 1.1 & 0.039 $\pm$ 0.003 $\pm$ 0.002 \\ 
 3.000 - 3.200 & 11.1 $\pm$ 1.2 $\pm$ 1.2 & 0.036 $\pm$ 0.002 $\pm$ 0.002 \\ 
 3.200 - 3.400 & 3.59 $\pm$ 0.48 $\pm$ 0.44 & 0.021 $\pm$ 0.001 $\pm$ 0.001 \\
 3.400 - 3.600 & 2.18 $\pm$ 0.31 $\pm$ 0.24 & 0.018 $\pm$ 0.001 $\pm$ 0.001 \\
 3.600 - 3.800 & 0.64 $\pm$ 0.25 $\pm$ 0.08 & 0.010 $\pm$ 0.002 $\pm$ 0.001 \\
\end{tabular}
\label{tabResEffESSI2}
\end{ruledtabular}
\end{center}
\end{table}

\section{\label{sec:FFratio} Proton FF ratio}
The proton FF ratio ${\rm R}$ is determined by fitting the distribution of the helicity angle $\theta_p$ for the selected data events. The helicity angle $\theta_p$ is the angle between the proton momentum in the $p \bar p $ rest frame, and the momentum of the $p \bar p   $ system in the $e^+e^-$ c.m. system. The distribution  of  $\theta_p$ is  given by \cite{Aubert:2005cb}

\be
\frac{dN}{d\cos\theta_p}=A(H_M(\cos\theta_p,M_{p \bar p })+ {\rm R}^2 H_E(\cos\theta_p,M_{p \bar p })),
\label{eqAngd}
\ee
where $A$ is an overall normalization parameter. The functions $H_M(\cos\theta_p,M_{p \bar p })$ and $H_E(\cos\theta_p,M_{p \bar p })$ describe the magnetic and the electric contributions to the angular distribution $\theta_p$, respectively. They are obtained from MC simulations in form of histograms. The process $e^+e^-\to p \bar p  \gamma$ is generated (up to the next to leading order radiative corrections) with $G_E=0$ to determine  $H_M$, and with $G_M=0$  to determine $H_E$.

\begin{figure}[h]
\includegraphics[height=6cm,width=8.5cm]{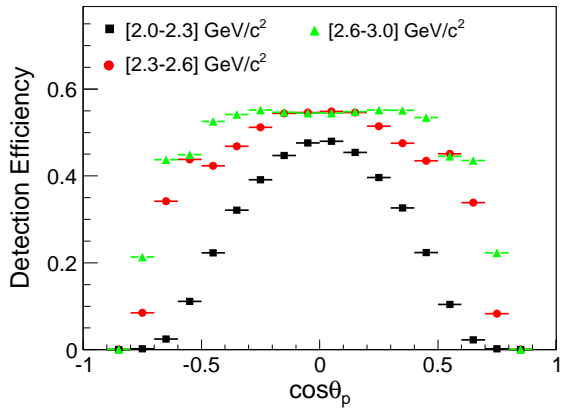}
\caption{Efficiency of the signal $e^+ e^- \to p \bar p  \gamma$  as a function of  $\cos\theta_p$   for different $M_{p\bar p }$ intervals at $\sqrt{s}=4.226$~GeV:  [2.0 - 2.3]~GeV/$c^2$  (black squares),  [2.3 - 2.6]~GeV/$c^2$  (red points),  and  [2.6 - 3.0]~GeV/$c^2$ (green triangles).}
\label{Effpolarangle4230}
\end{figure}

\begin{figure*}[th!]
\captionsetup[subfloat]{position=top,labelformat=empty}
  \subfloat[\label{costhetapatio1}]{\resizebox{0.34\textwidth}{!}{ \includegraphics{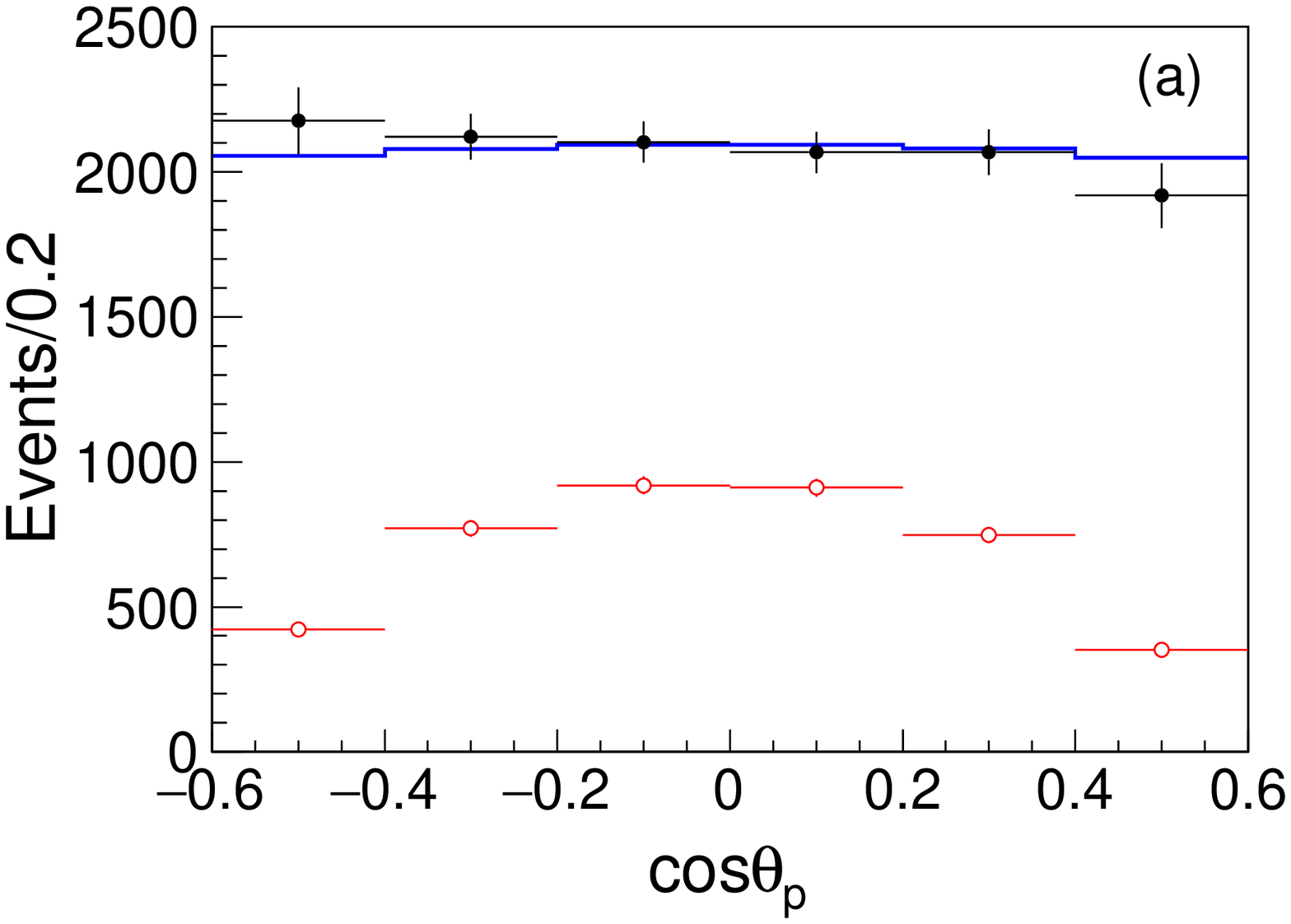} } }
  \subfloat[\label{costhetapatio2}]{\resizebox{0.34\textwidth}{!}{ \includegraphics{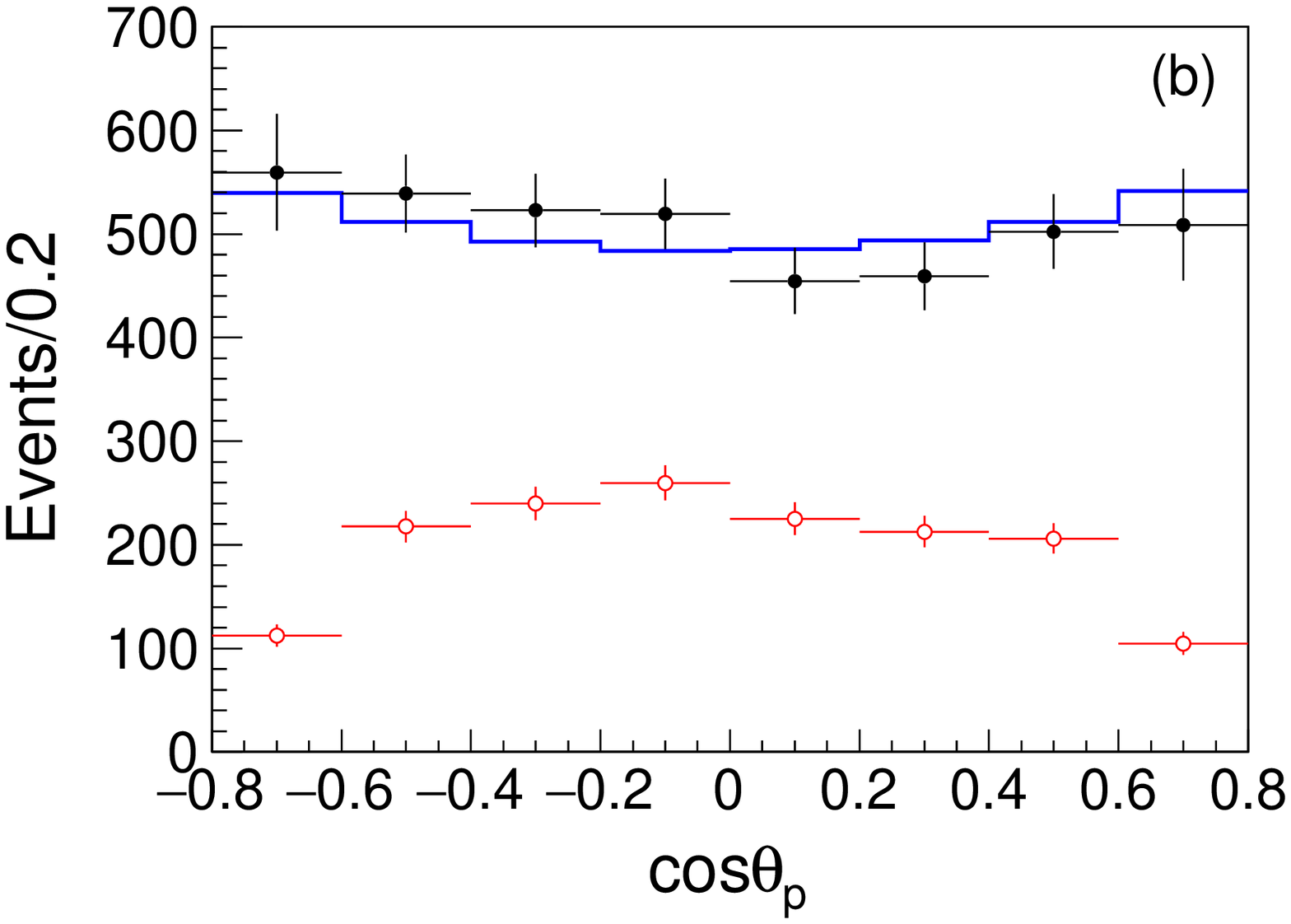} } }
  \subfloat[\label{costhetapatio3}]{\resizebox{0.34\textwidth}{!}{ \includegraphics{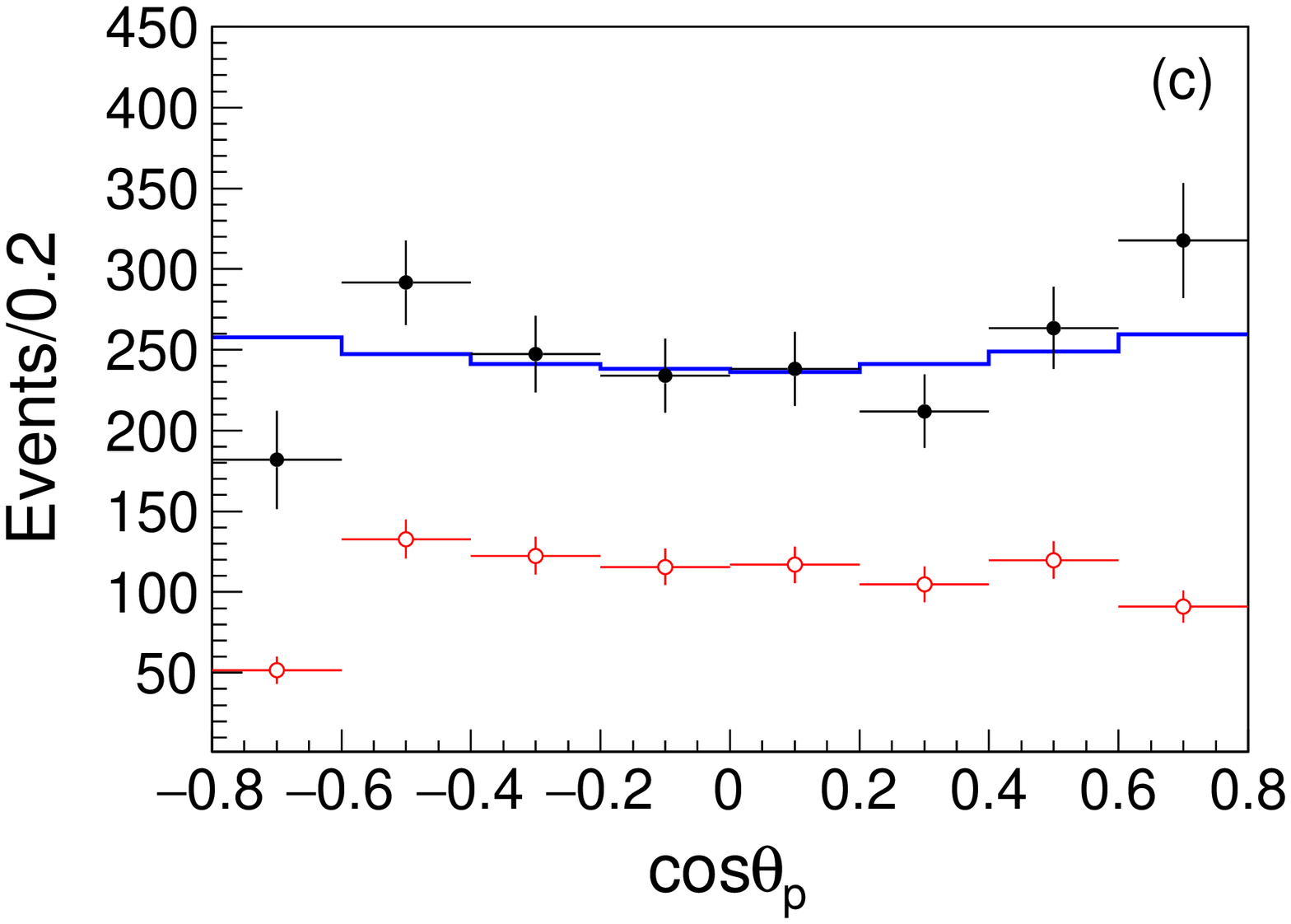} } }
\caption{Distributions of $\cos\theta_p$  for different $M_{p\bar p }$ intervals: (a) [2.0 - 2.3]~GeV/$c^2$, (b) [2.3 - 2.6]~GeV/$c^2$, and (c) [2.6 - 3.0]~GeV/$c^2$. The red open points with error bars represent the selected data events after background subtraction.  The black points are the data events for the signal channel corrected by the efficiency of the signal. The blue  histograms are the results of the fits.
}
  \label{costhetapatio}
 \end{figure*}

The angular distributions of the selected events are  studied in three $M_{p\bar p }$ intervals  between 2.0 and 3.0~GeV/$c^2$. The background events are subtracted from the selected data events in each   $\cos\theta_p$ interval.  After background subtraction, the data events are corrected by the efficiency of the signal. The signal efficiency is determined from  the MC simulations of the signal by dividing  the  number of selected  events by the number of generated events.  The signal efficiency depends on the distributions of $\theta_p$, $M_{p \bar p}$, and $\sqrt{s}$.  Figure~\ref{Effpolarangle4230} shows the distributions of the  signal efficiency as a function of $\cos\theta_p$  in the three $M_{p\bar p }$ intervals  at  $\sqrt{s}=4.226$~GeV. The data collected at the 7 c.m.  energies  are combined after efficiency correction.  The proton FF ratio is determined by fitting the $\cos\theta_p$ distributions  (Fig.~\ref{costhetapatio}) using   Eq.~(\ref{eqAngd}) and taking into account the relative normalization between $H_E$ and $H_M$.

\begin{figure}[h]
\includegraphics[height=7cm,width=9cm]{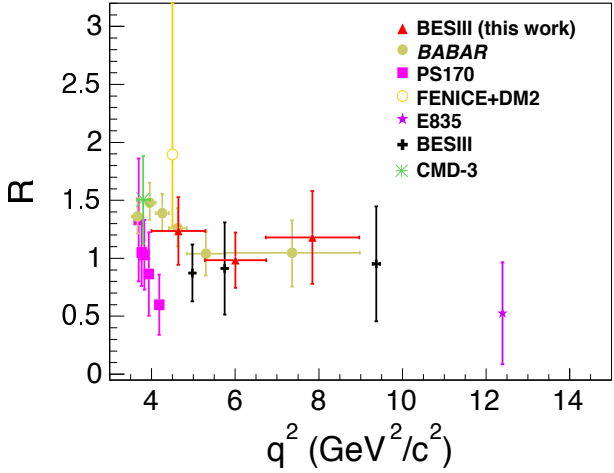}
\caption{Values of the proton FF ratio ${\rm R}$  measured in this analysis and in previous experiments: {\it BABAR}~\protect\cite{Lees:2013rqd,Lees:2013ebn}, PS170 (LEAR)~\protect\cite{Bardin:1994am}, BESIII~\protect\cite{Ablikim:2015vga}, CMD-3~\protect\cite{Akhmetshin:2015ifg}, and from Ref.~\protect\cite{Baldini:2005xx}. 
  The previous BESIII results (black crosses) were obtained using the energy scan technique where the precision on $q^2$ is given by the precise determination of $\sqrt{s}$.}
\label{ratiototpid}
\end{figure}

The obtained values of ${\rm R}$  are listed in Table~\ref{tabFFratiosys}. The total uncertainty is dominated by the statistical uncertainties.  The main contributions to the systematic uncertainty in the ${\rm R}$ measurements come from the fit range, background estimation, and from the $ M_{\rm miss}^2$ and $\theta_{\rm miss}$ requirements. A comparison of ${\rm R}$ measured in this work and other experiments is shown in Fig.~\ref{ratiototpid}.

\begin{table}[h]
\begin{center}
\caption{Measured  ${\rm R}$ (${\rm R}=|G_E|/|G_M|$) in each $M_{p\bar p }$ interval between 2.0 and 3.0~GeV/$c^2$. The quoted uncertainties are the sums of the statistical and systematic uncertainties in quadrature. The statistical uncertainties are dominant.}
\begin{ruledtabular}
\begin{tabular}{ccc}
$M_{p \bar p }$ [GeV/$c^2$] &   Fitting range ($\cos\theta_p$) & ${\rm R}$ \\ \hline
2.0 - 2.3   & [-0.6,0.6] &   $ 1.24 \pm 0.29$ \\
2.3 - 2.6   &[-0.8,0.8]  &   $0.98 \pm 0.24$\\
2.6 - 3.0   & [-0.8,0.8] &   $ 1.18 \pm 0.40$ \\
\end{tabular}
\label{tabFFratiosys}
\end{ruledtabular}
\end{center}
\end{table}

\section{\label{sec:branching} \boldmath{Branching fractions of $J/\psi,~\lowercase{\psi(3686) \to p \bar p}$ }}
The measured numbers of  resonance  decays $N_{{\cal{R}}}$ (${\cal{R}}=J/\psi,~\psi(3686)$)  (Sec.~\ref{sec:resdecay}) are used to determine the  branching fractions, $J/\psi \to p \bar p  $ and $\psi(3686) \to p \bar p  $, as follows~\cite{Benayoun:1999hm}:
\be
\Gamma_{{\cal{R}}\to e^+e^-} \times {\cal B}({{\cal{R}} \to p \bar p })=\frac{s M_{{\cal{R}}}}{12 \pi^2}\frac{N_{{\cal{R}}}}{ \epsilon_{{\cal{R}}} (1+\delta_{{\cal{R}}}) W(s,x_{{\cal{R}}}){\cal {L}}},
\ee
where $M_{{\cal{R}}}$ is the mass of the resonance,  $W(s,x_{{\cal{R}}})$ is the ISR  function [Eq.~(\ref{eq:isrc2})],   and $\Gamma_{{\cal{R}} \to e^+e^-}$  is the electronic width of  ${{\cal{R}}}$.  The radiative correction factor  $(1+\delta_{{\cal{R}}})$  is determined using the  MC events of the signal process $e^+ e^- \to p \bar p \gamma$.  The luminosity ${\cal {L}}$ is the integrated luminosity collected at the c.m. energy $\sqrt{s}$ (Table~\ref{dataset}). For the electronic widths of  $J/\psi$  and $\psi(3686)$, the nominal values from Ref. \cite{Patrignani:2016xqp} are used. MC samples for $J/\psi \to p \bar p  $ and $\psi(3686) \to p \bar p  $ are generated at the different c.m. energies between 3.773 and 4.6 GeV to determine the detection efficiency $\epsilon_{{\cal{R}}}$. The MC events are produced with  proton angular distributions described by the function $1+\mathcal C \cos^2\theta$  with $\mathcal C=0.595 \pm 0.012 \pm 0.015$ for $J/\psi$ \cite{Ablikim:2012eu} and $\mathcal C=1.03 \pm 0.06 \pm 0.03$ for $\psi(3686)$  \cite{Ablikim:2018ttl}. The branching fractions of $J/\psi \to p \bar p  $ and $\psi(3686) \to p \bar p  $ are calculated for each data sample individually.
The systematic uncertainties of  the measured  branching fractions include uncertainties from tracking ($2.0\%$), PID ($2.0\%$), $E_{\rm EMC}/p_{\rm rec}$  requirement ($1.0\%$), $M_{\rm miss}^2$ and $\theta_{\rm miss}$ requirements, luminosity determination (0.8$\%$), and radiator function $W(s,x)$ (0.5$\%$). The uncertainties from the $\theta_{\rm miss}$  ($M_{\rm miss}^2$) requirements are found to be $1.3\%$ ($1.0\%$) for  $\psi(3686)$ and negligible for $J/\psi$. The model error in the detection efficiency due to the uncertainty of the $\mathcal C$ value is negligible. The difference between the fit output using a linear and an exponential fit function for the nonpeaking events is added to the systematic uncertainties ($1.8\%$ for  $\psi(3686)$ and negligible for $J/\psi$).
The obtained average value of ${\cal B}({J/\psi \to p \bar p  })=(2.08 \pm  0.04 \pm 0.07) \times 10^{-3}$, where the quoted uncertainties are statistical and systematic, respectively, is in good agreement with the world average value  of $(2.12 \pm  0.03) \times 10^{-3}$ \cite{Patrignani:2016xqp}. For ${\cal B}({\psi(3686) \to p \bar p })$, the obtained average value $(3.01 \pm 0.23 \pm 0.12) \times 10^{-4}$  is consistent with  the world average value  of $(2.88  \pm 0.09) \times 10^{-4}$ \cite{Patrignani:2016xqp} and with the latest measurement of BESIII  ${\cal B}({\psi(3686) \to p \bar p })= (3.05  \pm 0.02 \pm 0.12) \times 10^{-4}$ \cite{Ablikim:2018ttl} based on $1.07 \times 10^{8}$ $\psi(3686)$  events \cite{Ablikim:2012pj}.

\section{\label{sec:conclusions} Summary}
Based on data samples corresponding to an integrated luminosity of 7.5 fb$^{-1}$ collected with the BESIII detector at c.m. energies between 3.773 and 4.600~GeV, the proton FFs have been measured using the  ISR technique.  In this work, the  $e^+ e^- \to p \bar p  \gamma$ events in which the ISR photons cannot be detected have been analyzed. The Born cross section of the $e^+ e^- \to p \bar p $ channel and the proton effective FF have been measured in 30 $M_{p\bar p }$ intervals between 2.0 and 3.8~GeV/$c^2$.  The results are consistent with previous measurements and provide better precision in different $M_{p\bar p }$ intervals. The total relative uncertainty of the Born cross section is between $8\%$ and $41\%$.  We have confirmed the structures seen in the  measurements of the proton effective FF by the {\it BABAR} Collaboration ~\cite{Lees:2013rqd,Lees:2013ebn}.
The proton angular distributions have been also analyzed to determine the proton FF ratio in 3 $M_{p\bar p }$  intervals between 2.0 and 3.0~GeV/$c^2$. The uncertainty on the measured proton FF ratio is dominated by the statistical uncertainty due to limited range of the proton angular distribution. The possibility to access the low $M_{p\bar p }$  region below 2~GeV/$c^2$ with ISR technique and undetected photon will be investigated in the future using the data samples collected at c.m. energies below 3.773~GeV.
In addition, the branching fractions of the $J/\psi,~\psi(3686)$ to $p \bar p  $ decays are also measured. The results are  in good agreements with the world average values.
BESIII is an excellent laboratory for the measurement of baryon timelike FFs. Both ISR and scan methods can be performed, and the kinematical threshold for different baryon pair production is covered by the energy range of BEPCII.  In 2015, BESIII performed high luminosity scan in 22 energy points between 2.0 and 3.08~GeV. Based on these data samples, more measurements of the nucleon  electromagnetic FFs will be available in this kinematical region.

\begin{acknowledgments}
The BESIII Collaboration thanks the staff of BEPCII and the IHEP computing center for their strong support. This work is supported in part by National Key Basic Research Program of China under Contract No. 2015CB856700; National Natural Science Foundation of China (NSFC) under Contracts No. 11335008, No. 11425524, No. 11625523, No. 11635010, No. 11735014; the Chinese Academy of Sciences (CAS) Large-Scale Scientific Facility Program; the CAS Center for Excellence in Particle Physics (CCEPP); Joint Large-Scale Scientific Facility Funds of the NSFC and CAS under Contracts No. U1532257, No. U1532258, No. U1732263; CAS Key Research Program of Frontier Sciences under Contracts No. QYZDJ-SSW-SLH003, No. QYZDJ-SSW-SLH040; 100 Talents Program of CAS; INPAC and Shanghai Key Laboratory for Particle Physics and Cosmology; German Research Foundation DFG under Contracts No. Collaborative Research Center CRC 1044, No. FOR 2359; Istituto Nazionale di Fisica Nucleare, Italy; Koninklijke Nederlandse Akademie van Wetenschappen (KNAW) under Contract No. 530-4CDP03; Ministry of Development of Turkey under Contract No. DPT2006K-120470; National Science and Technology fund; The Swedish Research Council; The Knut and Alice Wallenberg Foundation; U. S. Department of Energy under Contracts No. DE-FG02-05ER41374, No. DE-SC-0010118, No. DE-SC-0010504, No. DE-SC-0012069; University of Groningen (RuG) and the Helmholtzzentrum fuer Schwerionenforschung GmbH (GSI), Darmstadt.
\end{acknowledgments}

\bibliography{bibmain}

\begin{thebibliography}{53}%
\makeatletter
\providecommand \@ifxundefined [1]{%
 \@ifx{#1\undefined}
}%
\providecommand \@ifnum [1]{%
 \ifnum #1\expandafter \@firstoftwo
 \else \expandafter \@secondoftwo
 \fi
}%
\providecommand \@ifx [1]{%
 \ifx #1\expandafter \@firstoftwo
 \else \expandafter \@secondoftwo
 \fi
}%
\providecommand \natexlab [1]{#1}%
\providecommand \enquote  [1]{``#1''}%
\providecommand \bibnamefont  [1]{#1}%
\providecommand \bibfnamefont [1]{#1}%
\providecommand \citenamefont [1]{#1}%
\providecommand \href@noop [0]{\@secondoftwo}%
\providecommand \href [0]{\begingroup \@sanitize@url \@href}%
\providecommand \@href[1]{\@@startlink{#1}\@@href}%
\providecommand \@@href[1]{\endgroup#1\@@endlink}%
\providecommand \@sanitize@url [0]{\catcode `\\12\catcode `\$12\catcode
  `\&12\catcode `\#12\catcode `\^12\catcode `\_12\catcode `\%12\relax}%
\providecommand \@@startlink[1]{}%
\providecommand \@@endlink[0]{}%
\providecommand \url  [0]{\begingroup\@sanitize@url \@url }%
\providecommand \@url [1]{\endgroup\@href {#1}{\urlprefix }}%
\providecommand \urlprefix  [0]{URL }%
\providecommand \Eprint [0]{\href }%
\providecommand \doibase [0]{http://dx.doi.org/}%
\providecommand \selectlanguage [0]{\@gobble}%
\providecommand \bibinfo  [0]{\@secondoftwo}%
\providecommand \bibfield  [0]{\@secondoftwo}%
\providecommand \translation [1]{[#1]}%
\providecommand \BibitemOpen [0]{}%
\providecommand \bibitemStop [0]{}%
\providecommand \bibitemNoStop [0]{.\EOS\space}%
\providecommand \EOS [0]{\spacefactor3000\relax}%
\providecommand \BibitemShut  [1]{\csname bibitem#1\endcsname}%
\let\auto@bib@innerbib\@empty
\bibitem [{\citenamefont {Denig}\ and\ \citenamefont
  {Salm\`{e}}(2013)}]{Denig:2012by}%
  \BibitemOpen
  \bibfield  {author} {\bibinfo {author} {\bibfnamefont {A.}~\bibnamefont
  {Denig}}\ and\ \bibinfo {author} {\bibfnamefont {G.}~\bibnamefont
  {Salm\`{e}}},\ }\href@noop {\doibase 10.1016/j.ppnp.2012.09.005} {\bibfield
  {journal} {\bibinfo  {journal} {Prog. Part. Nucl. Phys.}\ }\textbf {\bibinfo
  {volume} {68}},\ \bibinfo {pages} {113} (\bibinfo {year} {2013})}\BibitemShut
  {NoStop}%
\bibitem [{\citenamefont {Pacetti}\ \emph {et~al.}(2015)\citenamefont
  {Pacetti}, \citenamefont {Baldini~Ferroli},\ and\ \citenamefont
  {Tomasi-Gustafsson}}]{Pacetti:2015iqa}%
  \BibitemOpen
  \bibfield  {author} {\bibinfo {author} {\bibfnamefont {S.}~\bibnamefont
  {Pacetti}}, \bibinfo {author} {\bibfnamefont {R.}~\bibnamefont
  {Baldini~Ferroli}}, \ and\ \bibinfo {author} {\bibfnamefont {E.}~\bibnamefont
  {Tomasi-Gustafsson}},\ }\href@noop {\doibase 10.1016/j.physrep.2014.09.005}
  {\bibfield  {journal} {\bibinfo  {journal} {Phys. Rep.}\ }\textbf {\bibinfo
  {volume} {550-551}},\ \bibinfo {pages} {1} (\bibinfo {year}
  {2015})}\BibitemShut {NoStop}%
\bibitem [{\citenamefont {Kuraev}\ \emph {et~al.}(2012)\citenamefont {Kuraev},
  \citenamefont {Tomasi-Gustafsson},\ and\ \citenamefont
  {Dbeyssi}}]{Kuraev:2011vq}%
  \BibitemOpen
  \bibfield  {author} {\bibinfo {author} {\bibfnamefont {E.~A.}\ \bibnamefont
  {Kuraev}}, \bibinfo {author} {\bibfnamefont {E.}~\bibnamefont
  {Tomasi-Gustafsson}}, \ and\ \bibinfo {author} {\bibfnamefont
  {A.}~\bibnamefont {Dbeyssi}},\ }\href@noop {\doibase
  10.1016/j.physletb.2012.04.073} {\bibfield  {journal} {\bibinfo  {journal}
  {Phys. Lett. B}\ }\textbf {\bibinfo {volume} {712}},\ \bibinfo {pages} {240}
  (\bibinfo {year} {2012})}\BibitemShut {NoStop}%
\bibitem [{\citenamefont {Akhiezer}\ and\ \citenamefont
  {Rekalo}(1968)}]{Akhiezer:1968ek}%
  \BibitemOpen
  \bibfield  {author} {\bibinfo {author} {\bibfnamefont {A.~I.}\ \bibnamefont
  {Akhiezer}}\ and\ \bibinfo {author} {\bibfnamefont {M.}~\bibnamefont
  {Rekalo}},\ }\href@noop {} {\bibfield  {journal} {\bibinfo  {journal} {Sov.
  Phys. Dokl.}\ }\textbf {\bibinfo {volume} {13}},\ \bibinfo {pages} {572}
  (\bibinfo {year} {1968})}\BibitemShut {NoStop}%
\bibitem [{\citenamefont {Akhiezer}\ and\ \citenamefont
  {Rekalo}(1974)}]{Akhiezer:1974em}%
  \BibitemOpen
  \bibfield  {author} {\bibinfo {author} {\bibfnamefont {A.~I.}\ \bibnamefont
  {Akhiezer}}\ and\ \bibinfo {author} {\bibfnamefont {M.}~\bibnamefont
  {Rekalo}},\ }\href@noop {} {\bibfield  {journal} {\bibinfo  {journal} {Sov.
  J. Part. Nucl.}\ }\textbf {\bibinfo {volume} {4}},\ \bibinfo {pages} {277}
  (\bibinfo {year} {1974})}\BibitemShut {NoStop}%
\bibitem [{\citenamefont {Puckett}\ \emph {et~al.}(2017)\citenamefont {Puckett}
  \emph {et~al.}}]{Puckett:2017flj}%
  \BibitemOpen
  \bibfield  {author} {\bibinfo {author} {\bibfnamefont {A.~J.~R.}\
  \bibnamefont {Puckett}} \emph {et~al.},\ }\href@noop {\doibase
  10.1103/PhysRevC.98.019907, 10.1103/PhysRevC.96.055203} {\bibfield  {journal}
  {\bibinfo  {journal} {Phys. Rev. C}\ }\textbf {\bibinfo {volume} {96}},\
  \bibinfo {pages} {055203} (\bibinfo {year} {2017})}\BibitemShut {NoStop}%
\bibitem [{\citenamefont {Castellano}\ \emph {et~al.}(1973)\citenamefont
  {Castellano}, \citenamefont {Giugno}, \citenamefont {Humphrey}, \citenamefont
  {Palmieri}, \citenamefont {Troise}, \citenamefont {Troya},\ and\
  \citenamefont {Vitale}}]{Castellano:1973}%
  \BibitemOpen
  \bibfield  {author} {\bibinfo {author} {\bibfnamefont {M.}~\bibnamefont
  {Castellano}}, \bibinfo {author} {\bibfnamefont {G.}~\bibnamefont {Giugno}},
  \bibinfo {author} {\bibfnamefont {J.}~\bibnamefont {Humphrey}}, \bibinfo
  {author} {\bibfnamefont {E.~S.}\ \bibnamefont {Palmieri}}, \bibinfo {author}
  {\bibfnamefont {G.}~\bibnamefont {Troise}}, \bibinfo {author} {\bibfnamefont
  {U.}~\bibnamefont {Troya}}, \ and\ \bibinfo {author} {\bibfnamefont
  {S.}~\bibnamefont {Vitale}},\ }\href@noop {\doibase 10.1007/BF02734600} {\bibfield
   {journal} {\bibinfo  {journal} {Nuovo Cimento A}\ }\textbf {\bibinfo
  {volume} {14}},\ \bibinfo {pages} {1} (\bibinfo {year} {1973})}\BibitemShut
  {NoStop}%
\bibitem [{\citenamefont {Andreotti}\ \emph {et~al.}(2003)\citenamefont
  {Andreotti} \emph {et~al.}}]{Andreotti:2003bt}%
  \BibitemOpen
  \bibfield  {author} {\bibinfo {author} {\bibfnamefont {M.}~\bibnamefont
  {Andreotti}} \emph {et~al.},\ }\href@noop {\doibase 10.1016/S0370-2693(03)00300-9}
  {\bibfield  {journal} {\bibinfo  {journal} {Phys. Lett. B}\ }\textbf
  {\bibinfo {volume} {559}},\ \bibinfo {pages} {20} (\bibinfo {year}
  {2003})}\BibitemShut {NoStop}%
\bibitem [{\citenamefont {Ambrogiani}\ \emph {et~al.}(1999)\citenamefont
  {Ambrogiani} \emph {et~al.}}]{Ambrogiani:1999bh}%
  \BibitemOpen
  \bibfield  {author} {\bibinfo {author} {\bibfnamefont {M.}~\bibnamefont
  {Ambrogiani}} \emph {et~al.} (\bibinfo {collaboration} {E835
  Collaboration}),\ }\href@noop {\doibase 10.1103/PhysRevD.60.032002} {\bibfield
  {journal} {\bibinfo  {journal} {Phys. Rev. D}\ }\textbf {\bibinfo {volume}
  {60}},\ \bibinfo {pages} {032002} (\bibinfo {year} {1999})}\BibitemShut
  {NoStop}%
\bibitem [{\citenamefont {Antonelli}\ \emph {et~al.}(1998)\citenamefont
  {Antonelli} \emph {et~al.}}]{Antonelli:1998fv}%
  \BibitemOpen
  \bibfield  {author} {\bibinfo {author} {\bibfnamefont {A.}~\bibnamefont
  {Antonelli}} \emph {et~al.},\ }\href@noop {\doibase 10.1016/S0550-3213(98)00083-2}
  {\bibfield  {journal} {\bibinfo  {journal} {Nucl. Phys. B}\ }\textbf
  {\bibinfo {volume} {517}},\ \bibinfo {pages} {3} (\bibinfo {year}
  {1998})}\BibitemShut {NoStop}%
\bibitem [{\citenamefont {Bardin}\ \emph {et~al.}(1994)\citenamefont {Bardin}
  \emph {et~al.}}]{Bardin:1994am}%
  \BibitemOpen
  \bibfield  {author} {\bibinfo {author} {\bibfnamefont {G.}~\bibnamefont
  {Bardin}} \emph {et~al.},\ }\href@noop {\doibase 10.1016/0550-3213(94)90052-3}
  {\bibfield  {journal} {\bibinfo  {journal} {Nucl. Phys. B}\ }\textbf
  {\bibinfo {volume} {411}},\ \bibinfo {pages} {3} (\bibinfo {year}
  {1994})}\BibitemShut {NoStop}%
\bibitem [{\citenamefont {Armstrong}\ \emph {et~al.}(1993)\citenamefont
  {Armstrong} \emph {et~al.}}]{Armstrong:1992wq}%
  \BibitemOpen
  \bibfield  {author} {\bibinfo {author} {\bibfnamefont {T.~A.}\ \bibnamefont
  {Armstrong}} \emph {et~al.} (\bibinfo {collaboration} {E760 Collaboration}),\
  }\href@noop {\doibase 10.1103/PhysRevLett.70.1212} {\bibfield  {journal} {\bibinfo
   {journal} {Phys. Rev. Lett.}\ }\textbf {\bibinfo {volume} {70}},\ \bibinfo
  {pages} {1212} (\bibinfo {year} {1993})}\BibitemShut {NoStop}%
\bibitem [{\citenamefont {Delcourt}\ \emph {et~al.}(1979)\citenamefont
  {Delcourt} \emph {et~al.}}]{Delcourt:1979ed}%
  \BibitemOpen
  \bibfield  {author} {\bibinfo {author} {\bibfnamefont {B.}~\bibnamefont
  {Delcourt}} \emph {et~al.},\ }\href@noop {\doibase 10.1016/0370-2693(79)90864-5}
  {\bibfield  {journal} {\bibinfo  {journal} {Phys. Lett. B}\ }\textbf
  {\bibinfo {volume} {86}},\ \bibinfo {pages} {395} (\bibinfo {year}
  {1979})}\BibitemShut {NoStop}%
\bibitem [{\citenamefont {Bisello}\ \emph {et~al.}(1983)\citenamefont {Bisello}
  \emph {et~al.}}]{Bisello:1983at}%
  \BibitemOpen
  \bibfield  {author} {\bibinfo {author} {\bibfnamefont {D.}~\bibnamefont
  {Bisello}} \emph {et~al.},\ }\href@noop {\doibase 10.1016/0550-3213(83)90381-4}
  {\bibfield  {journal} {\bibinfo  {journal} {Nucl. Phys.}\ }\textbf {\bibinfo
  {volume} {B224}},\ \bibinfo {pages} {379} (\bibinfo {year}
  {1983})}\BibitemShut {NoStop}%
\bibitem [{\citenamefont {Bisello}\ \emph {et~al.}(1990)\citenamefont {Bisello}
  \emph {et~al.}}]{Bisello:1990rf}%
  \BibitemOpen
  \bibfield  {author} {\bibinfo {author} {\bibfnamefont {D.}~\bibnamefont
  {Bisello}} \emph {et~al.} (\bibinfo {collaboration} {DM2 Collaboration}),\
  }\href@noop {\doibase 10.1007/BF01565602} {\bibfield  {journal} {\bibinfo
  {journal} {Z. Phys. C}\ }\textbf {\bibinfo {volume} {48}},\ \bibinfo {pages}
  {23} (\bibinfo {year} {1990})}\BibitemShut {NoStop}%
\bibitem [{\citenamefont {Ablikim}\ \emph {et~al.}(2005)\citenamefont {Ablikim}
  \emph {et~al.}}]{Ablikim:2005nn}%
  \BibitemOpen
  \bibfield  {author} {\bibinfo {author} {\bibfnamefont {M.}~\bibnamefont
  {Ablikim}} \emph {et~al.} (\bibinfo {collaboration} {BES Collaboration}),\
  }\href@noop {\doibase 10.1016/j.physletb.2005.09.044} {\bibfield  {journal}
  {\bibinfo  {journal} {Phys. Lett. B}\ }\textbf {\bibinfo {volume} {630}},\
  \bibinfo {pages} {14} (\bibinfo {year} {2005})}\BibitemShut {NoStop}%
\bibitem [{\citenamefont {Ablikim}\ \emph
  {et~al.}(2015{\natexlab{a}})\citenamefont {Ablikim} \emph
  {et~al.}}]{Ablikim:2015vga}%
  \BibitemOpen
  \bibfield  {author} {\bibinfo {author} {\bibfnamefont {M.}~\bibnamefont
  {Ablikim}} \emph {et~al.} (\bibinfo {collaboration} {BESIII Collaboration}),\
  }\href@noop {\doibase 10.1103/PhysRevD.91.112004} {\bibfield  {journal} {\bibinfo
  {journal} {Phys. Rev. D}\ }\textbf {\bibinfo {volume} {91}},\ \bibinfo
  {pages} {112004} (\bibinfo {year} {2015}{\natexlab{a}})}\BibitemShut
  {NoStop}%
\bibitem [{\citenamefont {Pedlar}\ \emph {et~al.}(2005)\citenamefont {Pedlar}
  \emph {et~al.}}]{Pedlar:2005sj}%
  \BibitemOpen
  \bibfield  {author} {\bibinfo {author} {\bibfnamefont {T.~K.}\ \bibnamefont
  {Pedlar}} \emph {et~al.} (\bibinfo {collaboration} {CLEO Collaboration}),\
  }\href@noop {\doibase 10.1103/PhysRevLett.95.261803} {\bibfield  {journal}
  {\bibinfo  {journal} {Phys. Rev. Lett.}\ }\textbf {\bibinfo {volume} {95}},\
  \bibinfo {pages} {261803} (\bibinfo {year} {2005})}\BibitemShut {NoStop}%
\bibitem [{\citenamefont {Akhmetshin}\ \emph {et~al.}(2016)\citenamefont
  {Akhmetshin} \emph {et~al.}}]{Akhmetshin:2015ifg}%
  \BibitemOpen
  \bibfield  {author} {\bibinfo {author} {\bibfnamefont {R.~R.}\ \bibnamefont
  {Akhmetshin}} \emph {et~al.} (\bibinfo {collaboration} {CMD-3
  Collaboration}),\ }\href@noop {\doibase 10.1016/j.physletb.2016.04.048} {\bibfield
   {journal} {\bibinfo  {journal} {Phys. Lett. B}\ }\textbf {\bibinfo {volume}
  {759}},\ \bibinfo {pages} {634} (\bibinfo {year} {2016})}\BibitemShut
  {NoStop}%
\bibitem [{\citenamefont {Lees}\ \emph
  {et~al.}(2013{\natexlab{a}})\citenamefont {Lees} \emph
  {et~al.}}]{Lees:2013rqd}%
  \BibitemOpen
  \bibfield  {author} {\bibinfo {author} {\bibfnamefont {J.~P.}\ \bibnamefont
  {Lees}} \emph {et~al.} (\bibinfo {collaboration} {{\it BABAR}
  Collaboration}),\ }\href@noop {\doibase 10.1103/PhysRevD.88.032011} {\bibfield
  {journal} {\bibinfo  {journal} {Phys. Rev. D}\ }\textbf {\bibinfo {volume}
  {88}},\ \bibinfo {pages} {032011} (\bibinfo {year}
  {2013}{\natexlab{a}})}\BibitemShut {NoStop}%
\bibitem [{\citenamefont {Lees}\ \emph
  {et~al.}(2013{\natexlab{b}})\citenamefont {Lees} \emph
  {et~al.}}]{Lees:2013ebn}%
  \BibitemOpen
  \bibfield  {author} {\bibinfo {author} {\bibfnamefont {J.~P.}\ \bibnamefont
  {Lees}} \emph {et~al.} (\bibinfo {collaboration} {{\it BABAR}
  Collaboration}),\ }\href@noop {\doibase 10.1103/PhysRevD.87.092005} {\bibfield
  {journal} {\bibinfo  {journal} {Phys. Rev. D}\ }\textbf {\bibinfo {volume}
  {87}},\ \bibinfo {pages} {092005} (\bibinfo {year}
  {2013}{\natexlab{b}})}\BibitemShut {NoStop}%
\bibitem [{\citenamefont {Lorentz}\ \emph {et~al.}(2015)\citenamefont
  {Lorentz}, \citenamefont {Hammer},\ and\ \citenamefont
  {Meissner}}]{Lorenz:2015pba}%
  \BibitemOpen
  \bibfield  {author} {\bibinfo {author} {\bibfnamefont {I.~T.}\ \bibnamefont
  {Lorentz}}, \bibinfo {author} {\bibfnamefont {H.-W.}\ \bibnamefont {Hammer}},
  \ and\ \bibinfo {author} {\bibfnamefont {U.-G.}\ \bibnamefont {Meissner}},\
  }\href@noop {\doibase doi:10.1103/PhysRevD.92.034018} {\bibfield  {journal}
  {\bibinfo  {journal} {Phys. Rev. D}\ }\textbf {\bibinfo {volume} {92}},\
  \bibinfo {pages} {034018} (\bibinfo {year} {2015})}\BibitemShut {NoStop}%
\bibitem [{\citenamefont {Bianconi}\ and\ \citenamefont
  {Tomasi-Gustafsson}(2016)}]{Bianconi:2015vva}%
  \BibitemOpen
  \bibfield  {author} {\bibinfo {author} {\bibfnamefont {A.}~\bibnamefont
  {Bianconi}}\ and\ \bibinfo {author} {\bibfnamefont {E.}~\bibnamefont
  {Tomasi-Gustafsson}},\ }\href@noop {\doibase doi:10.1103/PhysRevC.93.035201}
  {\bibfield  {journal} {\bibinfo  {journal} {Phys. Rev. C}\ }\textbf {\bibinfo
  {volume} {93}},\ \bibinfo {pages} {035201} (\bibinfo {year}
  {2016})}\BibitemShut {NoStop}%
\bibitem [{\citenamefont {Ablikim}\ \emph {et~al.}(2010)\citenamefont {Ablikim}
  \emph {et~al.}}]{Ablikim:2009aa}%
  \BibitemOpen
  \bibfield  {author} {\bibinfo {author} {\bibfnamefont {M.}~\bibnamefont
  {Ablikim}} \emph {et~al.} (\bibinfo {collaboration} {BESIII Collaboration}),\
  }\href@noop {\doibase 10.1016/j.nima.2009.12.050} {\bibfield  {journal} {\bibinfo
  {journal} {Nucl. Instrum. Methods Phys. Res., Sect. A}\ }\textbf {\bibinfo
  {volume} {614}},\ \bibinfo {pages} {345} (\bibinfo {year}
  {2010})}\BibitemShut {NoStop}%
\bibitem [{\citenamefont {Druzhinin}\ \emph {et~al.}(2011)\citenamefont
  {Druzhinin}, \citenamefont {Eidelman}, \citenamefont {Serednyakov},\ and\
  \citenamefont {Solodov}}]{Druzhinin:2011qd}%
  \BibitemOpen
  \bibfield  {author} {\bibinfo {author} {\bibfnamefont {V.~P.}\ \bibnamefont
  {Druzhinin}}, \bibinfo {author} {\bibfnamefont {S.~I.}\ \bibnamefont
  {Eidelman}}, \bibinfo {author} {\bibfnamefont {S.~I.}\ \bibnamefont
  {Serednyakov}}, \ and\ \bibinfo {author} {\bibfnamefont {E.~P.}\ \bibnamefont
  {Solodov}},\ }\href@noop {\doibase 10.1103/RevModPhys.83.1545} {\bibfield
  {journal} {\bibinfo  {journal} {Rev. Mod. Phys}\ }\textbf {\bibinfo {volume}
  {83}},\ \bibinfo {pages} {1545} (\bibinfo {year} {2011})}\BibitemShut
  {NoStop}%
\bibitem [{\citenamefont {Tzara}(1970)}]{Tzara:1970ne}%
  \BibitemOpen
  \bibfield  {author} {\bibinfo {author} {\bibfnamefont {C.}~\bibnamefont
  {Tzara}},\ }\href@noop {\doibase 10.1016/0550-3213(70)90290-7} {\bibfield
  {journal} {\bibinfo  {journal} {Nucl. Phys.}\ }\textbf {\bibinfo {volume}
  {B18}},\ \bibinfo {pages} {246} (\bibinfo {year} {1970})}\BibitemShut
  {NoStop}%
\bibitem [{\citenamefont {Ablikim}\ \emph {et~al.}(2016)\citenamefont {Ablikim}
  \emph {et~al.}}]{Ablikim:2015orh}%
  \BibitemOpen
  \bibfield  {author} {\bibinfo {author} {\bibfnamefont {M.}~\bibnamefont
  {Ablikim}} \emph {et~al.} (\bibinfo {collaboration} {BESIII Collaboration}),\
  }\href@noop {\doibase 10.1016/j.physletb.2015.11.043} {\bibfield  {journal}
  {\bibinfo  {journal} {Phys. Lett. B}\ }\textbf {\bibinfo {volume} {753}},\
  \bibinfo {pages} {629} (\bibinfo {year} {2016})}\BibitemShut {NoStop}%
\bibitem [{\citenamefont {Ablikim}\ \emph
  {et~al.}(2015{\natexlab{b}})\citenamefont {Ablikim} \emph
  {et~al.}}]{Ablikim:2015nan}%
  \BibitemOpen
  \bibfield  {author} {\bibinfo {author} {\bibfnamefont {M.}~\bibnamefont
  {Ablikim}} \emph {et~al.} (\bibinfo {collaboration} {BESIII Collaboration}),\
  }\href@noop {\doibase 10.1088/1674-1137/39/9/093001} {\bibfield  {journal}
  {\bibinfo  {journal} {Chin. Phys. C}\ }\textbf {\bibinfo {volume} {39}},\
  \bibinfo {pages} {093001} (\bibinfo {year} {2015}{\natexlab{b}})}\BibitemShut
  {NoStop}%
\bibitem [{\citenamefont {Agostinelli}\ \emph {et~al.}(2003)\citenamefont
  {Agostinelli} \emph {et~al.}}]{Agostinelli:2002hh}%
  \BibitemOpen
  \bibfield  {author} {\bibinfo {author} {\bibfnamefont {S.}~\bibnamefont
  {Agostinelli}} \emph {et~al.} (\bibinfo {collaboration} {GEANT4
  Collaboration}),\ }\href@noop {\doibase 10.1016/S0168-9002(03)01368-8} {\bibfield
  {journal} {\bibinfo  {journal} {Nucl. Instrum. Methods Phys. Res., Sect. A}\
  }\textbf {\bibinfo {volume} {1308}},\ \bibinfo {pages} {250} (\bibinfo {year}
  {2003})}\BibitemShut {NoStop}%
\bibitem [{\citenamefont {Deng}\ \emph {et~al.}(2006)\citenamefont {Deng} \emph
  {et~al.}}]{Deng:2006}%
  \BibitemOpen
  \bibfield  {author} {\bibinfo {author} {\bibfnamefont {Z.~Y.}\ \bibnamefont
  {Deng}} \emph {et~al.},\ }\href@noop {} {\bibfield  {journal} {\bibinfo
  {journal} {High Energy Phys. Nucl. Phys.}\ }\textbf {\bibinfo {volume}
  {30}},\ \bibinfo {pages} {371} (\bibinfo {year} {2006})}\BibitemShut
  {NoStop}%
\bibitem [{\citenamefont {Czyz}\ \emph {et~al.}(2014)\citenamefont {Czyz},
  \citenamefont {Kuehn},\ and\ \citenamefont {Tracz}}]{Czyz:2014sha}%
  \BibitemOpen
  \bibfield  {author} {\bibinfo {author} {\bibfnamefont {H.}~\bibnamefont
  {Czyz}}, \bibinfo {author} {\bibfnamefont {J.~H.}\ \bibnamefont {Kuehn}}, \
  and\ \bibinfo {author} {\bibfnamefont {S.}~\bibnamefont {Tracz}},\ }\href@noop
  {\doibase 10.1103/PhysRevD.90.114021} {\bibfield  {journal} {\bibinfo
  {journal} {Phys. Rev. D}\ }\textbf {\bibinfo {volume} {90}},\ \bibinfo
  {pages} {114021} (\bibinfo {year} {2014})}\BibitemShut {NoStop}%
\bibitem [{\citenamefont {Sjostrand}()}]{Sjostrand:1994ek}%
  \BibitemOpen
  \bibfield  {author} {\bibinfo {author} {\bibfnamefont {T.}~\bibnamefont
  {Sjostrand}},\ }\href@noop {} {\ }\href@noop
  {http://arxiv.org/abs/hep-ph/9508391} {arXiv:hep-ph/9508391} \BibitemShut
  {NoStop}%
\bibitem [{\citenamefont {Ping}\ \emph {et~al.}(2014)\citenamefont {Ping} \emph
  {et~al.}}]{Ping:2013jka}%
  \BibitemOpen
  \bibfield  {author} {\bibinfo {author} {\bibfnamefont {R.~G.}\ \bibnamefont
  {Ping}} \emph {et~al.},\ }\href@noop {\doibase 10.1088/1674-1137/38/8/083001}
  {\bibfield  {journal} {\bibinfo  {journal} {Chin. Phys. C}\ }\textbf
  {\bibinfo {volume} {38}},\ \bibinfo {pages} {083001} (\bibinfo {year}
  {2014})}\BibitemShut {NoStop}%
\bibitem [{\citenamefont {Ping}(2008)}]{Ping:2008zz}%
  \BibitemOpen
  \bibfield  {author} {\bibinfo {author} {\bibfnamefont {R.~G.}\ \bibnamefont
  {Ping}},\ }\href@noop {\doibase 10.1088/1674-1137/32/8/001} {\bibfield  {journal}
  {\bibinfo  {journal} {Chin. Phys. C}\ }\textbf {\bibinfo {volume} {32}},\
  \bibinfo {pages} {599} (\bibinfo {year} {2008})}\BibitemShut {NoStop}%
\bibitem [{\citenamefont {Jadach}\ \emph {et~al.}(2001)\citenamefont {Jadach},
  \citenamefont {Ward},\ and\ \citenamefont {Was}}]{Jadach:2000ir}%
  \BibitemOpen
  \bibfield  {author} {\bibinfo {author} {\bibfnamefont {S.}~\bibnamefont
  {Jadach}}, \bibinfo {author} {\bibfnamefont {B.~F.~L.}\ \bibnamefont {Ward}},
  \ and\ \bibinfo {author} {\bibfnamefont {Z.}~\bibnamefont {Was}},\ }\href@noop
  {\doibase 10.1103/PhysRevD.63.113009} {\bibfield  {journal} {\bibinfo
  {journal} {Phys. Rev. D}\ }\textbf {\bibinfo {volume} {63}},\ \bibinfo
  {pages} {113009} (\bibinfo {year} {2001})}\BibitemShut {NoStop}%
\bibitem [{\citenamefont {Jadach}\ \emph {et~al.}(2000)\citenamefont {Jadach},
  \citenamefont {Ward},\ and\ \citenamefont {Was}}]{Jadach:1999vf}%
  \BibitemOpen
  \bibfield  {author} {\bibinfo {author} {\bibfnamefont {S.}~\bibnamefont
  {Jadach}}, \bibinfo {author} {\bibfnamefont {B.~F.~L.}\ \bibnamefont {Ward}},
  \ and\ \bibinfo {author} {\bibfnamefont {Z.}~\bibnamefont {Was}},\ }\href@noop
  {\doibase 10.1016/S0010-4655(00)00048-5} {\bibfield  {journal} {\bibinfo
  {journal} {Comput. Phys. Commun.}\ }\textbf {\bibinfo {volume} {130}},\
  \bibinfo {pages} {260} (\bibinfo {year} {2000})}\BibitemShut {NoStop}%
\bibitem [{\citenamefont {Balossini}\ \emph {et~al.}(2006)\citenamefont
  {Balossini}, \citenamefont {Carloni~Calame}, \citenamefont {Montagna},
  \citenamefont {Nicrosini},\ and\ \citenamefont
  {Piccinini}}]{Balossini:2006wc}%
  \BibitemOpen
  \bibfield  {author} {\bibinfo {author} {\bibfnamefont {G.}~\bibnamefont
  {Balossini}}, \bibinfo {author} {\bibfnamefont {C.~M.}\ \bibnamefont
  {Carloni~Calame}}, \bibinfo {author} {\bibfnamefont {G.}~\bibnamefont
  {Montagna}}, \bibinfo {author} {\bibfnamefont {O.}~\bibnamefont {Nicrosini}},
  \ and\ \bibinfo {author} {\bibfnamefont {F.}~\bibnamefont {Piccinini}},\
  }\href@noop {\doibase 10.1016/j.nuclphysb.2006.09.022} {\bibfield  {journal}
  {\bibinfo  {journal} {Nucl. Phys. B}\ }\textbf {\bibinfo {volume} {758}},\
  \bibinfo {pages} {227} (\bibinfo {year} {2006})}\BibitemShut {NoStop}%
\bibitem [{\citenamefont {Bonneau}\ and\ \citenamefont
  {Martin}(1971)}]{Bonneau:1971mk}%
  \BibitemOpen
  \bibfield  {author} {\bibinfo {author} {\bibfnamefont {G.}~\bibnamefont
  {Bonneau}}\ and\ \bibinfo {author} {\bibfnamefont {F.}~\bibnamefont
  {Martin}},\ }\href@noop {\doibase 10.1016/0550-3213(71)90102-7} {\bibfield
  {journal} {\bibinfo  {journal} {Nucl. Phys.}\ }\textbf {\bibinfo {volume}
  {B27}},\ \bibinfo {pages} {381} (\bibinfo {year} {1971})}\BibitemShut
  {NoStop}%
\bibitem [{\citenamefont {Lange}(2001)}]{Lange:2001uf}%
  \BibitemOpen
  \bibfield  {author} {\bibinfo {author} {\bibfnamefont {D.~J.}\ \bibnamefont
  {Lange}},\ }\href@noop {\doibase 10.1016/S0168-9002(01)00089-4} {\bibfield
  {journal} {\bibinfo  {journal} {Nucl. Instrum. Methods Phys. Res., Sect. A}\
  }\textbf {\bibinfo {volume} {462}},\ \bibinfo {pages} {152} (\bibinfo {year}
  {2001})}\BibitemShut {NoStop}%
\bibitem [{\citenamefont {Ablikim}\ \emph {et~al.}(2014)\citenamefont {Ablikim}
  \emph {et~al.}}]{Ablikim:2014kxa}%
  \BibitemOpen
  \bibfield  {author} {\bibinfo {author} {\bibfnamefont {M.}~\bibnamefont
  {Ablikim}} \emph {et~al.} (\bibinfo {collaboration} {BESIII Collaboration}),\
  }\href@noop {\doibase 10.1103/PhysRevD.90.032007} {\bibfield  {journal} {\bibinfo
  {journal} {Phys. Rev. D}\ }\textbf {\bibinfo {volume} {90}},\ \bibinfo
  {pages} {032007} (\bibinfo {year} {2014})}\BibitemShut {NoStop}%
\bibitem [{\citenamefont {Ablikim}\ \emph {et~al.}(2017)\citenamefont {Ablikim}
  \emph {et~al.}}]{Ablikim:2017gtb}%
  \BibitemOpen
  \bibfield  {author} {\bibinfo {author} {\bibfnamefont {M.}~\bibnamefont
  {Ablikim}} \emph {et~al.} (\bibinfo {collaboration} {BESIII Collaboration}),\
  }\href@noop {\doibase 10.1016/j.physletb.2017.05.033} {\bibfield  {journal}
  {\bibinfo  {journal} {Phys. Lett. B}\ }\textbf {\bibinfo {volume} {771}},\
  \bibinfo {pages} {45} (\bibinfo {year} {2017})}\BibitemShut {NoStop}%
\bibitem [{\citenamefont {Schmelling}(1995)}]{1402}%
  \BibitemOpen
  \bibfield  {author} {\bibinfo {author} {\bibfnamefont {M.}~\bibnamefont
  {Schmelling}},\ }\href@noop {http://stacks.iop.org/1402-4896/51/i=6/a=002}
  {\bibfield  {journal} {\bibinfo  {journal} {Phys. Scr.}\ }\textbf {\bibinfo
  {volume} {51}},\ \bibinfo {pages} {676} (\bibinfo {year} {1995})}\BibitemShut
  {NoStop}%
\bibitem [{\citenamefont {Brodsky}\ and\ \citenamefont
  {de~Teramond}(2008)}]{Brodsky:2007hb}%
  \BibitemOpen
  \bibfield  {author} {\bibinfo {author} {\bibfnamefont {S.~J.}\ \bibnamefont
  {Brodsky}}\ and\ \bibinfo {author} {\bibfnamefont {G.~F.}\ \bibnamefont
  {de~Teramond}},\ }\href@noop {\doibase 10.1103/PhysRevD.77.056007} {\bibfield
  {journal} {\bibinfo  {journal} {Phys. Rev. D}\ }\textbf {\bibinfo {volume}
  {77}},\ \bibinfo {pages} {056007} (\bibinfo {year} {2008})}\BibitemShut
  {NoStop}%
\bibitem [{\citenamefont {Tomasi-Gustafsson}\ and\ \citenamefont
  {Rekalo}(2001)}]{TomasiGustafsson:2001za}%
  \BibitemOpen
  \bibfield  {author} {\bibinfo {author} {\bibfnamefont {E.}~\bibnamefont
  {Tomasi-Gustafsson}}\ and\ \bibinfo {author} {\bibfnamefont {M.~P.}\
  \bibnamefont {Rekalo}},\ }\href@noop {\doibase 10.1016/S0370-2693(01)00312-4}
  {\bibfield  {journal} {\bibinfo  {journal} {Phys. Lett. B}\ }\textbf
  {\bibinfo {volume} {504}},\ \bibinfo {pages} {291} (\bibinfo {year}
  {2001})}\BibitemShut {NoStop}%
\bibitem [{\citenamefont {Shirkov}\ and\ \citenamefont
  {Solovtsov}(1997)}]{Shirkov:1997wi}%
  \BibitemOpen
  \bibfield  {author} {\bibinfo {author} {\bibfnamefont {D.~V.}\ \bibnamefont
  {Shirkov}}\ and\ \bibinfo {author} {\bibfnamefont {I.~L.}\ \bibnamefont
  {Solovtsov}},\ }\href@noop {\doibase 10.1103/PhysRevLett.79.1209} {\bibfield
  {journal} {\bibinfo  {journal} {Phys. Rev. Lett.}\ }\textbf {\bibinfo
  {volume} {79}},\ \bibinfo {pages} {1209} (\bibinfo {year}
  {1997})}\BibitemShut {NoStop}%
\bibitem [{\citenamefont {Bianconi}\ and\ \citenamefont
  {Tomasi-Gustafsson}(2015)}]{Bianconi:2015owa}%
  \BibitemOpen
  \bibfield  {author} {\bibinfo {author} {\bibfnamefont {A.}~\bibnamefont
  {Bianconi}}\ and\ \bibinfo {author} {\bibfnamefont {E.}~\bibnamefont
  {Tomasi-Gustafsson}},\ }\href@noop {\doibase 10.1103/PhysRevLett.114.232301}
  {\bibfield  {journal} {\bibinfo  {journal} {Phys. Rev. Lett.}\ }\textbf
  {\bibinfo {volume} {114}},\ \bibinfo {pages} {232301} (\bibinfo {year}
  {2015})}\BibitemShut {NoStop}%
\bibitem [{\citenamefont {Patrignani}\ \emph {et~al.}(2016)\citenamefont
  {Patrignani} \emph {et~al.}}]{Patrignani:2016xqp}%
  \BibitemOpen
  \bibfield  {author} {\bibinfo {author} {\bibfnamefont {C.}~\bibnamefont
  {Patrignani}} \emph {et~al.} (\bibinfo {collaboration} {Particle Data
  Group}),\ }\href@noop {\doibase 10.1088/1674-1137/40/10/100001} {\bibfield
  {journal} {\bibinfo  {journal} {Chin. Phys. C}\ }\textbf {\bibinfo {volume}
  {40}},\ \bibinfo {pages} {100001} (\bibinfo {year} {2016})}\BibitemShut
  {NoStop}%
\bibitem [{\citenamefont {Aubert}\ \emph {et~al.}(2006)\citenamefont {Aubert}
  \emph {et~al.}}]{Aubert:2005cb}%
  \BibitemOpen
  \bibfield  {author} {\bibinfo {author} {\bibfnamefont {B.}~\bibnamefont
  {Aubert}} \emph {et~al.} (\bibinfo {collaboration} {{\it BABAR}
  Collaboration}),\ }\href@noop {\doibase 10.1103/PhysRevD.73.012005} {\bibfield
  {journal} {\bibinfo  {journal} {Phys. Rev. D}\ }\textbf {\bibinfo {volume}
  {73}},\ \bibinfo {pages} {012005} (\bibinfo {year} {2006})}\BibitemShut
  {NoStop}%
\bibitem [{\citenamefont {Baldini}\ \emph {et~al.}(2006)\citenamefont {Baldini}
  \emph {et~al.}}]{Baldini:2005xx}%
  \BibitemOpen
  \bibfield  {author} {\bibinfo {author} {\bibfnamefont {R.}~\bibnamefont
  {Baldini}} \emph {et~al.},\ }\href@noop {\doibase 10.1140/epjc/s2006-02499-4}
  {\bibfield  {journal} {\bibinfo  {journal} {Eur. Phys. J. C}\ }\textbf
  {\bibinfo {volume} {46}},\ \bibinfo {pages} {421} (\bibinfo {year}
  {2006})}\BibitemShut {NoStop}%
\bibitem [{\citenamefont {Benayoun}\ \emph {et~al.}(1999)\citenamefont
  {Benayoun}, \citenamefont {Eidelman}, \citenamefont {Ivanchenko},\ and\
  \citenamefont {Silagadze}}]{Benayoun:1999hm}%
  \BibitemOpen
  \bibfield  {author} {\bibinfo {author} {\bibfnamefont {M.}~\bibnamefont
  {Benayoun}}, \bibinfo {author} {\bibfnamefont {S.~I.}\ \bibnamefont
  {Eidelman}}, \bibinfo {author} {\bibfnamefont {V.~N.}\ \bibnamefont
  {Ivanchenko}}, \ and\ \bibinfo {author} {\bibfnamefont {Z.~K.}\ \bibnamefont
  {Silagadze}},\ }\href@noop {\doibase 10.1142/S021773239900273X} {\bibfield
  {journal} {\bibinfo  {journal} {Mod. Phys. Lett. A}\ }\textbf {\bibinfo
  {volume} {14}},\ \bibinfo {pages} {2605} (\bibinfo {year}
  {1999})}\BibitemShut {NoStop}%
\bibitem [{\citenamefont {Ablikim}\ \emph {et~al.}(2012)\citenamefont {Ablikim}
  \emph {et~al.}}]{Ablikim:2012eu}%
  \BibitemOpen
  \bibfield  {author} {\bibinfo {author} {\bibfnamefont {M.}~\bibnamefont
  {Ablikim}} \emph {et~al.} (\bibinfo {collaboration} {BESIII Collaboration}),\
  }\href@noop {\doibase 10.1103/PhysRevD.86.032014} {\bibfield  {journal} {\bibinfo
  {journal} {Phys. Rev. D}\ }\textbf {\bibinfo {volume} {86}},\ \bibinfo
  {pages} {032014} (\bibinfo {year} {2012})}\BibitemShut {NoStop}%
\bibitem [{\citenamefont {Ablikim}\ \emph {et~al.}(2018)\citenamefont {Ablikim}
  \emph {et~al.}}]{Ablikim:2018ttl}%
  \BibitemOpen
  \bibfield  {author} {\bibinfo {author} {\bibfnamefont {M.}~\bibnamefont
  {Ablikim}} \emph {et~al.} (\bibinfo {collaboration} {BESIII Collaboration}),\
  }\href@noop {\doibase 10.1103/PhysRevD.98.032006} {\bibfield  {journal} {\bibinfo
  {journal} {Phys. Rev. D}\ }\textbf {\bibinfo {volume} {98}},\ \bibinfo
  {pages} {032006} (\bibinfo {year} {2018})}\BibitemShut {NoStop}%
\bibitem [{\citenamefont {Ablikim}\ \emph {et~al.}(2013)\citenamefont {Ablikim}
  \emph {et~al.}}]{Ablikim:2012pj}%
  \BibitemOpen
  \bibfield  {author} {\bibinfo {author} {\bibfnamefont {M.}~\bibnamefont
  {Ablikim}} \emph {et~al.} (\bibinfo {collaboration} {BESIII Collaboration}),\
  }\href@noop {\doibase 10.1088/1674-1137/37/6/063001} {\bibfield  {journal}
  {\bibinfo  {journal} {Chin. Phys. C}\ }\textbf {\bibinfo {volume} {37}},\
  \bibinfo {pages} {063001} (\bibinfo {year} {2013})}\BibitemShut {NoStop}%
\end{thebibliography}%

\end{document}